\documentclass[12pt,draftclsnofoot,onecolumn]{IEEEtran}
\usepackage{amsfonts,dsfont,amssymb,amsbsy,amsmath,paralist,theorem,bm,ifthen,color}
\usepackage[pdfstartview=FitH,bookmarksnumbered,unicode,bookmarksopen=true]{hyperref}
\usepackage{graphicx}
\usepackage{epstopdf}
\usepackage{multirow}
\usepackage{subfigure,relsize}
\usepackage{mathrsfs}
\usepackage{bm}
\usepackage{stfloats}
\usepackage{placeins}
\usepackage{url}
\usepackage[noadjust]{cite}
\usepackage{cases,booktabs}
\usepackage{graphicx,algorithmic,algorithm,relsize}
\usepackage[justification=centering]{caption} \usepackage[table]{xcolor}

\newtheorem{Proposition}{Proposition}
\newtheorem{Theorem}{Theorem}

\newcommand\Cc{\ensuremath{\mathcal{C}}}

\newcommand\Sc{\ensuremath{\mathcal{S}}}

\newcommand\Lc{\ensuremath{{\mathcal{L}}}}

\newcommand\Mc{\ensuremath{{\mathcal{M}}}}
\newcommand\Uc{\ensuremath{{\mathcal{U}}}}

\newcommand\wb{\ensuremath{{\bf w}}}

\newcommand\ub{\ensuremath{{\bm u}}}

\newcommand\hb{\ensuremath{{\bf h}}}
\newcommand\Ab{\ensuremath{{\bf A}}}

\newcommand\cb{\ensuremath{{\bf c}}}

\newcommand\Db{\ensuremath{{\bf D}}}

\newcommand\Ic{\ensuremath{{\mathcal{I}}}}

\newcommand\xb{\ensuremath{{\bf x}}}

\newcommand\qb{\ensuremath{{\bf q}}}

\newcommand\vb{\ensuremath{{\bf v}}}
\newcommand\Wb{\ensuremath{{\bf W}}}

\newcommand\Bc{\ensuremath{{\mathcal{B}  }}}
\newcommand\Dc{\ensuremath{{\mathcal{D}  }}}
\newcommand\Qc{\ensuremath{{\mathcal{Q}}}}

\newcommand\thetab{\ensuremath{{\bm \theta}}}

\newcommand\zerob{\ensuremath{{\bm 0}}}

\newcommand{\wt}{\widetilde}

\newcommand{\ul}{\underline}

\DeclareMathOperator{\trace}{Tr}

\newcommand\E{\ensuremath{{\mathbb{E}}}}

\newcommand\diag{\ensuremath{{\rm diag}}}

\newcommand\Rbb{\ensuremath{{\mathbb{R}}}}

\newcommand\Xc{\ensuremath{{\mathcal{X}}}}

\graphicspath{{fig/}}

\definecolor{green}{RGB}{34	195	46}
\definecolor{red}{RGB}{220 0 0}

\newcommand\Zbb{\ensuremath{{\mathbb{Z}}}}

\usepackage{graphicx} \allowdisplaybreaks[4]

\title{Communication-Learning Co-Design for Pinching-Antenna-Assisted Federated Learning in Heterogeneous Wireless Networks}

\author{Shuai Wang, Wenxuan Sun, 
Mingjie Shao, \\
Wantao Jiang, 
Yanqing Xu,
Lingxiang~Li, 
and Zhi Chen
     \thanks{\smaller[1] S. Wang, W. Jiang, L. Li, and Z. Chen are with the National Key Laboratory of Wireless Communications, University of Electronic Science and Technology of China, Chengdu 611731, China (email: shuaiwang@uestc.edu.cn, 202522220241@std.uestc.edu.cn, lingxiang.li@uestc.edu.cn, chenzhi@uestc.edu.cn).}
     \thanks{\smaller[1] W. Sun is with the School of Information Science and Engineering, Shandong University, Qingdao 266200, China (email: wenxuansun@mail.sdu.edu.cn).}   
 \thanks{\smaller[1] M. Shao is with the State Key Laboratory of Mathematical Sciences, AMSS, Chinese Academy of Science, Beijing 100080, China (email: mingjieshao@amss.ac.cn).}
       \thanks{\smaller[1] Y. Xu is with the School of Electronic and Information Engineering, South China University of Technology, Guangzhou 510641, China (email: xuyanqing@scut.edu.cn).}
}
	
\vspace{-0.3cm}

\date{\today}

\begin{document}

\maketitle

\begingroup
\linespread{1.2}\selectfont
\begin{abstract}

Wireless federated learning (FL) suffers from two coupled bottlenecks: statistical heterogeneity across clients and communication-constrained model synchronization over unreliable wireless links. Instead of focusing on algorithmic design and/or wireless resource optimization, this paper provides a systematic study of pinching antenna (PA)-empowered FL for heterogeneous wireless networks and develops a unified communication-and-learning framework. Specifically, we integrate a segmented waveguide-enabled pinching antenna (SWAN) architecture with a heterogeneity-robust FL algorithm based on client variance reduction and dual quantization, so as to jointly mitigate communication bottlenecks and client drift. On this basis, we further develop a two-stage SWAN-assisted downlink protocol and formulate fairness-oriented uplink optimization problems for PA positioning, bandwidth allocation, and quantization design. Efficient solution frameworks are then derived for the resulting downlink and uplink optimization problems. Experimental results show the superiority of the proposed design in convergence speed and final accuracy under non-independent and identically distributed data and communication bottlenecks.

\end{abstract}

\endgroup

 \vspace{-0.1cm}
\section{Introduction} 

Federated learning (FL) has emerged as a promising paradigm for enabling distributed intelligence at the network edge, thereby supporting the increasing demand for network intelligence in sixth-generation (6G) wireless systems \cite{FedVRA_Shuai_2023}. In FL, a central server coordinates a large number of mobile clients to collaboratively train a shared model through iterative local computation and global aggregation, without requiring the exchange of raw data \cite{li2025differentially}. Despite its promise, FL over wireless networks faces two fundamental challenges: 1) clients typically possess non-independent and identically distributed (non-i.i.d.) data, leading to significant statistical data heterogeneity; 2) client-server model synchronization relies on wireless links that are often resource-constrained, unreliable, and error-prone. These two issues jointly hinder training efficiency and lead to significant model performance degradation, substantially limiting the scalability and practical deployment of FL in wireless environments \cite{FL_MoblieEdge_2019}.

To address data heterogeneity, a wide range of FL algorithms have been developed, most of which build upon FedAvg by introducing mechanisms such as local regularization, adaptive aggregation, momentum-based updates, and variance reduction \cite{FedQVR_Shuai_2025}. In essence, they correct local or global update directions so as to better align them with the global objective, thereby alleviating the adverse impact of non-i.i.d. data. However, most focus on algorithmic optimization and implicitly assume ideal or sufficiently reliable communication conditions. Such an assumption is often unrealistic in practical wireless networks, where limited communication resources, channel fading, transmission errors, and intermittent connectivity may severely impair client-server model exchange. Although several studies reduce communication overhead through model compression, device scheduling, and wireless resource optimization, the resulting performance gains remain fundamentally constrained by the intrinsic limitations of the wireless channel \cite{wang2025robust}\cite{wang2022quantized}.

To fundamentally alleviate the communication bottleneck in wireless FL, it is necessary to move beyond conventional communication-efficient algorithm design and exploit physical-layer technologies that can actively reshape the propagation environment. In this context, pinching antennas (PAs), as a representative flexible-antenna technology, have recently emerged as a promising solution for wireless channel enhancement \cite{xu2025rate}. By leveraging dielectric waveguides together with dynamically activatable or movable pinching elements, PAs can establish flexible radiation points closer to users, thereby enabling effective link reconstruction and line-of-sight (LoS) restoration under blockage conditions \cite{xu2026attenuation}. Compared with conventional flexible-antenna schemes, such as fluid antennas and movable antennas, PAs support such reconfiguration over a much larger spatial range, making them particularly suitable for challenging propagation scenarios with severe path loss and LoS blockage. This capability, combined with low deployment cost, high flexibility, and strong scalability, makes PAs highly attractive for wireless FL, where reliable client-server communication is crucial for efficient and robust training.

Despite the above potential, the integration of PAs into FL systems remains largely unexplored. A few studies have considered PAs alone but overlooked data heterogeneity across clients \cite{Straggler-FL-pinn2026}\cite{e28030341}.  Different from conventional wireless optimization problems, FL performance couples communication quality with heterogeneous client dynamics. Thus, designing PA-assisted FL requires a unified framework jointly addressing link enhancement, communication resource allocation, and learning convergence.

\vspace{-0.4cm}
\subsection{Contributions}

In this paper, we investigate PA-empowered FL in heterogeneous wireless networks. In particular, we aim to jointly address the two coupled bottlenecks in wireless FL under PA-enabled communication. To this end, we integrate a segmented waveguide-enabled pinching antenna (SWAN) architecture with a heterogeneity-robust FL framework, and develop the corresponding downlink and uplink transmission designs. The contributions  can be elaborated from four aspects.

\begin{enumerate}
    \item We propose a SWAN-assisted heterogeneity-robust FL framework for wireless edge networks. Specifically, a SWAN architecture is deployed at the server side to enhance client-server communications, while a client-variance-reduced FL algorithm is developed to mitigate client drift under non-i.i.d. data. In addition, a dual-quantization mechanism is introduced to reduce  downlink and uplink communication overhead, where common quantization is adopted for downlink delivery and client-adaptive quantization is used for uplink transmission. Thus, the proposed framework unifies communication enhancement and heterogeneity-robust FL design.
    \item We develop a two-stage downlink protocol for FL, consisting of a global-model multicast stage and a control-variable unicast stage. The design jointly optimizes beamforming, PA positioning, time allocation, and quantization bits under Rician statistical channel state information (CSI) and a common latency budget. To tackle the resulting mixed-integer non-convex problem, we establish a rate-oriented solution framework, under which the original design is reduced to multicast and unicast rate optimization, followed by closed-form recovery of the optimal time allocation and downlink quantization level. This yields an efficient and structured downlink design for supportable delivery. 

    \item For uplink transmission, we develop a fairness-oriented SWAN-assisted design that optimizes uplink quantization, bandwidth allocation, and PA positions to maximize the minimum supportable quantization level across active clients. The novelty lies in uncovering the problem structure and deriving efficient solution frameworks for a hierarchy of uplink designs with increasing generality, including exact structural characterization of antenna-position optimization, and an equivalent inverse reformulation for the general joint design. These provide a systematic way to coordinate communication resources and SWAN geometry for robust uplink model delivery.

    \item We extensively evaluate the proposed design on the CIFAR-10 dataset under non-i.i.d. distributions and practical communication constraints. Experimental results show that the proposed method consistently and significantly outperforms representative baselines under various quantization settings and heterogeneity levels. Moreover, compared with a conventional base-station-based communication architecture, the proposed SWAN-assisted design achieves significant additional performance gains under communication bottlenecks, further validating the effectiveness of integrating PAs with heterogeneity-robust FL.
\end{enumerate}

{\bf Synopsis:} Section II presents the system model and preliminaries. Section III introduces the proposed heterogeneity-robust FL framework. Sections IV and V develop the SWAN-assisted downlink and uplink communication designs, respectively. Section VI presents the experimental results, and Section VII concludes the paper.

\vspace{-0.2cm}
\section{System Model and Preliminaries}
We consider a FL system in a wireless edge network covering a rectangular service area of size $L \times W$. For ease of exposition, the network geometry is described in a three dimensional Cartesian coordinate system. The edge server, typically co-located with a base station (BS), is positioned at $\qb^{\rm BS}=(0,0,H)$. Meanwhile, a set of $N$ clients, each equipped with a single antenna, are randomly distributed within the service area. Assume all clients lie on the same horizontal plane, the position of client $i, \forall i \in [N]$, is denoted by $\qb_i^c=(x_i,y_i,0)$, where $x_i\in[0,L], y_i\in[0,W]$.

\vspace{-0.3cm}
\subsection{Federated Learning}

Assume that each client $i$ processes a local dataset $\Dc_i$, the FL process aims to train an AI model $\thetab$ such that it performs the best across all local datasets $\{\Dc_i\}_{i = 1}^{N}$. Given a sample-wise loss function $f(\thetab; \xi)$ which measures the prediction error of  model $\thetab$ on any data sample $\xi$, the objective of FL is formulated as
\begin{align} \label{eqn: FL_prob}
	\min_{\substack{\thetab \in \Rbb^d}}~&  F(\thetab) \triangleq \frac{1}{N}\sum_{i = 1}^{N} F_i(\thetab; \Dc_i) \triangleq \E_{\xi \in \Dc_i} [f(\thetab; \xi)], 
\end{align}where $F_i(\thetab; \Dc_i)$ denotes the local loss function of client $i$, defined as the expected loss over its local dataset. Note  that, in practical wireless edge networks, clients typically collect and own data that reflects their individual contexts, leading to non-i.i.d. local datasets $\Dc_i, i \in [N]$ with sizes $\{n_i\}_{i=1}^{N}$.

Most FL algorithms solve problem \eqref{eqn: FL_prob} in an iterative model-averaging manner, with FedAvg being the most widely used method in wireless edge networks \cite{FedAvg_noniid_2019}. 
Specifically, at round $r$ of FedAvg, the server samples a fixed-size subset of clients $\Sc^r \subseteq [N]$ ($|\Sc^r| = P$) and broadcasts the current global model $\thetab^r$. Each selected client $i \in \Sc^r$ then performs $\tau$ steps of local stochastic gradient descent (SGD) on its dataset $\Dc_i$, i.e., $\forall t = 0, \ldots, \tau - 1$,
\begin{align}
	\thetab_i^{r, t+1} = 	\thetab_i^{r, t} - \eta g(\thetab_i^{r,t}),  \thetab_i^{r, 0} = \thetab^r, 
\end{align} where $\eta>0$ is the step size; $g(\thetab_i^{r,t}) \triangleq \frac{1}{|\Bc|}\sum_{ \xi \in \Bc}\nabla f(\thetab_i^{r,t}; \xi) $ is stochastic gradient (SG); $\Bc$ is a mini-batch uniformly sampled from $\Dc_i$. This yields an updated local model $\thetab_i^{r+1} = \thetab_i^{r, \tau}$. Upon completion of local training, the active clients upload their model updates $\{\thetab_i^{r+1} - \thetab^r\}_{i \in \Sc^r}$to the server through wireless links.  The server then aggregates the received updates to generate the global model, i.e., $\thetab^{r+1} = \thetab^{r} + \frac{1}{|\Sc^r|}\sum_{i \in \Sc^r}(\thetab_i^{r+1}-\thetab^{r}) $, which is subsequently broadcast back to the clients via wireless links for the next training round.

It is worth noting that the above training procedure is severely hindered by intrinsic data heterogeneity across clients \cite{FedAvg_noniid_2019}. In particular, non-i.i.d. data misaligns local and global objectives, resulting in significant inter-device variance during global aggregation. This variance leads to client drift, where local models deviate from the global optimum, ultimately causing slower convergence, possible divergence, and substantial performance degradation.

Meanwhile, wireless links between the server and clients are often unreliable, time-varying, and constrained by limited resources. Large server-client distances, complex propagation environments, and large AI model sizes further induce severe channel fading and a higher risk of LoS blockage. These impairments lead to excessive communication overhead, long delays, frequent transmission errors, and even link outages under stringent latency constraints, thereby disrupting the FL process and degrading learning performance. Therefore, jointly addressing data heterogeneity and communication bottlenecks is essential for efficient FL over wireless networks.

\vspace{-0.2cm}
\subsection{Segmented Waveguide Enabled PA}

To cope with adverse propagation conditions and severe LoS blockage, we employ a SWAN architecture at the server side to assist client-server communications \cite{SWAN-2026}. Specifically, a dielectric waveguide is deployed from the server (i.e., the base station) along the $x$-axis, i.e., on the line   $y=0$, at a fixed height $H$, on which $M$ PAs are mounted. By activating the PAs on the waveguide, additional radiation points can be flexibly generated to establish reliable wireless links between the server and distributed clients. 

Unlike a single-long waveguides, the SWAN architecture partitions the waveguide into $M$ contiguous segments with boundaries $0=l_0 < l_1 < \dots < l_M = L$, where $\Mc\triangleq\{1,2,\dots,M\}$. To ensure unique client association, define $\Ic_m=[l_{m-1},l_m)$ for $m<M$ and $\Ic_M=[l_{M-1},l_M]$. The PA on segment $m$ may move over the closed feasible interval $\Lc_m=[l_{m-1},l_m]$. The server is equipped with a multi-radio-frequency (multi-RF) feeding structure, where each radio-frequency (RF) chain connects to one segment \cite{SWAN-2026}. This segmented multi-RF design enables independent signal control across segments, thereby providing additional degrees of freedom for both transmission enhancement and user access. Each segment features a movable PA positioned at $\qb_m^{\rm PA}=(x_m^{\rm PA},0,H)$, where $x_m^{\mathrm{PA}}\in\Lc_m$. The horizontal coordinate $x_m^{\mathrm{PA}}$ can be continuously adjusted within its segment to adapt the propagation geometry to the client locations and channel conditions, thereby improving link quality and communication reliability.

We adopt a Rician fading channel model to characterize the cascaded propagation consisting of in-waveguide transmission and the free-space multi-path propagation from the PA to the client. Specifically, for segment $m$, its left boundary $l_{m-1}$ is assumed as the feed point, i.e., the point where the transmit signal from the server is coupled into the corresponding waveguide segment. Then, the channel between client $i$ and the PA on segment $m$ is modeled as
\begin{align} \label{eq:general_channel}
 h_{i,m}(\qb_i^c, x_m^{\mathrm{PA}}) &=  e^{-j\frac{2\pi}{\lambda_g}(x_m^{\mathrm{PA}} - l_{m-1})} \frac{\sqrt{\beta_0}}{d_{i,m}} \bigg( \sqrt{\frac{K_R}{K_R+1}} e^{-j\frac{2\pi}{\lambda}d_{i,m}} + \sqrt{\frac{1}{K_R+1}} z_{i,m} \bigg).
\end{align}
where $\beta_0$ denotes the reference path-loss gain; $\lambda$ is the carrier wavelength; $\lambda_g$ is the guided wavelength within the dielectric waveguide; $K_R$ denotes the Rician factor; $z_{i,m} \sim \mathcal{CN}(0, 1)$ represents the scattered small-scale fading component, whose realizations are assumed to be independent across clients and segments; $d_{i,m}$ is the physical distance between the PA on segment $m$ and client $i$, which is given by
\begin{align}
    d_{i,m}=\|\qb_i^c-\qb_m^{\rm PA}\|_2 = \sqrt{(x_i - x_m^{\mathrm{PA}})^2 + y_i^2 +H^2}.
\end{align}

\vspace{-0.3cm}
\section{Heterogeneity-robust FL Framework}
In this section, we present a heterogeneity-robust FL framework for the considered SWAN-assisted system. Built upon the model-averaging principle, the proposed framework incorporates a robust client-variance-reduction mechanism to mitigate the adverse effect of data heterogeneity and a dual quantization scheme to reduce the communication burden on FL training.

\vspace{-0.3cm}
\subsection{Client-variance-reduced FL}

Following client-variance-reduction in \cite{FedQVR_Shuai_2025}, we introduce two kinds of control variables to the vanilla FedAvg algorithm, namely, a global control variable $\cb^r$ and a local one $\cb_i^r$ for each client $i$, both maintained and updated at the server side. These control variables are designed to track the global and local update directions, respectively, and are used to correct the local descent direction, thereby mitigating client drift caused by the mismatch between local objectives and the global objective under data heterogeneity.

The detailed FL framework is summarized as follows. At round $r$, the server randomly samples a small subset of active clients $\Sc^r$ and broadcast the current global model $\thetab^r$ and local control variable $\cb_i^r$ to each client in $\Sc^r$. Then, each selected client performs a fixed number of local update steps over its private dataset $\Dc_i$ to obtain an updated local model $\thetab_i^{r+1}$. In particular, the local model of client $i, i \in \Sc^r$ is updated by perturbed SGD, i.e., $\forall t = 0, \ldots, \tau-1$,
\begin{align}		
	&\thetab_i^{r, t+1} = \frac{1}{1 + \gamma \eta} (\thetab_i^{r,t} - \eta(g_i(\thetab_i^{r,t})- \cb_i^r)) +  \frac{ \gamma \eta}{1 + \gamma \eta} \thetab^r, \label{eqn: thetai_update} 
\end{align}where $\thetab_i^{r, 0} = \thetab^r, \thetab_i^{r+1}=\thetab_i^{r,\tau}$;$\eta > 0$ is the local step size; $\gamma >0$ is a predefined constant; $\tau$ is the number of local update steps. The resulting model update $\thetab_i^{r+1} - \thetab^{r}$ is uploaded to the server. 
Upon receiving the model updates from all active clients, the server first forms the averaged intermediate model
\begin{align}
    \bar \thetab^{r+1} = \thetab^r + \frac{1}{|\Sc^r|} \sum_{i \in \Sc^r}(\thetab_i^{r+1} - \thetab^{r}).
\end{align}
\vspace{-0.3cm}

Based on $\bar \thetab^{r+1}$, the server then updates the local control variables for all clients as
\begin{align}\label{eqn: control_var_i}
	\cb_i^{r+1} =& \begin{cases}\cb_i^{r} -\frac{a_i}{\eta \wt \tau}(\thetab_i^{r+1} - \thetab^{r}), ~{\rm if} ~i \in \Sc^r, \\
		\cb_i^{r} - \frac{a_i}{\eta \wt \tau} (\bar \thetab^{r+1} - \thetab^{r}), ~{\rm otherwise},
	\end{cases}
\end{align}
where $\wt \tau \triangleq \frac{1}{\gamma\eta}\big(1- \frac{1}{(1 + \gamma\eta)^{\tau}}\big)$;$a_i
\in (0, 1)$ is a hyperparameter. After that, the global control variable is updated by averaging the local control variables across all clients, namely,
\begin{align}\label{eqn: control_var}
    \cb^{r+1} =& \frac{1}{N}\sum_{i=1}^{N}\cb_i^{r+1}.
\end{align}
Finally, the server updates the global model by combining the averaged model update with the correction term induced by the global control variable, i.e.,
\begin{align} \label{eqn: global_model}
	\thetab^{r+1} =&~ \bar \thetab^{r+1} - \frac{1}{\gamma} \cb^{r+1}.
\end{align}

The introduced control variables serve to align the local descent directions with the global update trajectory. In particular, for each active client $i \in \Sc^r$, the update of its local control variate can be rewritten as
\begin{align}
	\cb_i^{r+1} =  (1 - a_i)\cb_i^r + a_i \sum_{t= 0}^{\tau - 1}\frac{b_i^{r, t} }{\sum_{t= 0}^{\tau - 1}b_i^{r, t}} g_i(\thetab_i^{r,t}), 
\end{align}where $b_i^{r, t} =(\frac{1}{1 + \gamma \eta})^{\tau - t}$. This expression shows that  $\cb_i^r$ acts as an estimate of the local gradient direction by exponentially accumulating historical normalized local gradients. Since the global control variable $\cb^{r+1} = \frac{1}{N}\sum_{i=1}^{N} \cb_i^{r+1}$, it can be interpreted as an estimate of the global update direction obtained by averaging the local control variables across all clients. Furthermore, substituting the global model update \eqref{eqn: global_model} into the local iteration \eqref{eqn: thetai_update} yields
\begin{align}		
	\thetab_i^{r, t+1} =& \frac{\thetab_i^{r,t} - \eta(g_i(\thetab_i^{r,t})- \cb_i^r + \cb^r)}{1 + \gamma \eta}  +  \frac{ \gamma \eta \bar \thetab^{r}}{1 + \gamma \eta}. \label{eqn: cov_comb}  
\end{align}

As seen from \eqref{eqn: cov_comb}, by embedding both $\cb_i^r$ and $\cb^r$, the effective local update direction $g_i(\thetab_i^{r,t}) + \cb^r - \cb_i^{r}$ becomes much better aligned with the global one, thereby compensating for the mismatch between local and global objectives.  Moreover, the resulting convex combination with the averaged local models further stabilizes the local updates and prevents them from deviating excessively from the global trajectory. In this way, each client update is regularized toward the global descent direction through variance reduction across clients. As a result, the client drift issue can be significantly suppressed, which in turn accelerates the convergence of the global model under non-i.i.d. data.

\vspace{-0.3cm}

\subsection{Dual Quantization for Communication }

To relieve the communication burden, we introduce a dual-quantization mechanism, which, unlike conventional FL frameworks that focus solely on the uplink quantization, is applied to both downlink and uplink transmissions. The need for downlink quantization arises from the fact that the server sends not only a shared global model but also client-specific control variables in each round, which doubles the downlink communication cost.  Consequently, although the server is typically more communication-capable than clients, the resulting downlink traffic can still be substantial, thereby challenging efficient and reliable model delivery over wireless links. 

We follow the stochastic quantization method $\Qc(\ub, B)$ from \cite{wang2022quantized} to implement the proposed mechanism, where $\ub$ is the input model to be quantized and $B \geq 1$ is the number of quantization bits. The unbiased quantizer $\Qc(\cdot, B)$ compresses the input model into a low-bit representation.
The total number of communication bits needed for $\Qc(\cdot, B)$ is $d(B + 1) + \mu$, where $\mu$ is used to represent the upper and lower bounds of the input model, and thus a small $B$ can yield significant communication cost reduction per round.

Based on the above quantizer, the proposed dual-quantization mechanism adopts a common number of quantization bits across clients for downlink transmission and client-adaptive numbers of quantization bits for uplink transmission. In particular, since the downlink messages are sent by the same server and serve as the common training references for all active clients, they are quantized using the same number of quantization bits $B_D^r$, i.e., $\Qc(\thetab^r, B_D^r)$ and $\Qc(\cb_i^r, B_D^r)$, to maintain the algorithmic stability across clients. By contrast, the local update of each active client $i$ is quantized using an individual quantization bit number $B_i^r$, i.e., $\Qc(\thetab_i^{r+1} - \thetab^r, B_i^r)$, so as to adapt to the heterogeneous uplink channels and transmission capabilities of different clients.

Algorithm \ref{alg: FedAVR} presents the whole process. In each round, the server also determines the uplink transmission configuration and notifies the active clients through downlink control signaling. It is worth noting that both the downlink and uplink transmissions are tightly coupled with the SWAN-assisted communication system. Although model quantization with low bits can substantially reduce the communication overhead and facilitates transmission, it introduces severe quantization distortion, which adversely affects model aggregation and training convergence. Therefore, it is essential to fully exploit SWAN to balance communication efficiency and learning performance through effective transmission and resource allocation design. Such schemes correspond to Steps $5$-$7$ and $11$ of Algorithm \ref{alg: FedAVR}, which are presented in detail in subsequent sections.

\vspace{-0.1cm}
	\begin{algorithm}[bh!]
		\caption{Heterogeneity-robust FL framework}
		\label{alg: FedAVR}
		\begin{algorithmic}[1]
			\STATE {\bfseries Input:} initial values of $\thetab_i^0 = \thetab^0$, $\cb_i^0 = \cb^0=\zerob,  \eta > 0, \gamma > 0, 1\leq \tau \in \Zbb, 1 > a_i > 0, B_i^r, B_D^r \geq 1, \forall i, r$.
			\FOR{round $r=0$ {\bfseries to} $R - 1$}
			\STATE {\bfseries \ul{Server side:}} 
            \STATE Sample a set of clients $\Sc^r$ randomly from $[N]$.
            \STATE Compute bandwidth and quantization allocation $\{W_i^r,B_i^r\}_{i\in\Sc^r}$, and the uplink PA positions of SWAN via Algorithm 3 or 4.
            \STATE Design the downlink transmission via Algorithm 2.
            \STATE Deliver $\Qc(\thetab^r,B_D^r)$, $\Qc(\cb_i^r,B_D^r)$, and $\{W_i^r,B_i^r\}$ to the active clients.
			\STATE {\bfseries \ul{Client side:}}
			\FOR{client $i \in \Sc^r$ in parallel}
            \STATE Compute $\thetab_i^{r+1}$ via $\tau$ steps of SGD in \eqref{eqn: thetai_update}
			\STATE Compute and upload $\Delta_i^{r+1}=\Qc(\thetab_i^{r+1}-\thetab^r,B_i^r)$ over SWAN with the assigned uplink resource.
			\ENDFOR
			\STATE {\bfseries \ul{Server side:}} 
            \STATE Compute $\cb_i^{r+1}, \cb^{r+1}, \thetab^{r+1}$  via \eqref{eqn: control_var_i}, \eqref{eqn: control_var}, \eqref{eqn: global_model}, respectively.
			\ENDFOR
			\STATE {\bfseries Output:} $\thetab^R$.
		\end{algorithmic}
	\end{algorithm}

\vspace{-0.1cm}
\section{SWAN-ASSISTED DOWNLINK DESIGN}

In the downlink phase of each FL round, the server needs to deliver two types of quantized information to the scheduled clients: the common global model and the client-specific local control variables. To match this hybrid traffic structure, we adopt a two-stage downlink transmission protocol over SWAN, namely, a global-model multicast stage, and a control-variable unicast stage.

\begingroup
\emergencystretch=1em
In the global-model multicast stage, the server multicasts the quantized global model $\Qc(\thetab^r,B_D^r)$ to all active clients using the segment-multiplexing (SM) protocol, under which all segments simultaneously transmit the same information-bearing signal using a beamforming weight vector, thereby helping improve the worst-case received signal quality and the global-model delivery capability \cite{1634819}.  After that, the server delivers the quantized local control variables and two scalars $\{\Qc(\cb_i^r,B_D^r), W_i^r, B_i^r\}_{i\in\Sc^r}$ through time-division multiple access (TDMA)-based cooperative unicast, where each active client is served in an individual time slot. During the time slot assigned to client $i$, all segments cooperatively transmit $\{\Qc(\cb_i^r,B_D^r), W_i^r, B_i^r\}$ to this client \cite{5594708}. Since only one client is scheduled at a time, this stage is interference-free and reduces to a single-user downlink transmission.\par
\endgroup

On this basis, we formulate a two-stage downlink transmission design problem and propose a rate-oriented optimization framework for multicast and unicast beamforming, quantization bits, and PA positioning.

\vspace{-0.3cm}
\subsection{Downlink Signal Model}

For the downlink transmission, we adopt a metric based on the average signal-to-noise ratio (SNR) to characterize the long-term transmission capability under Rician statistical CSI \cite{xu2025random_los_nlos_pa}. Specifically, by averaging over the random non-line-of-sight (NLoS) scattering components in \eqref{eq:general_channel}, the average received SNR can be expressed through a Rician statistical channel correlation matrix. The deterministic LoS mean between segment $m$ and client $i$ is given by

\vspace{-0.3cm}
\begin{align}
\bar{h}_{i,m}(x_m^{\mathrm{PA}})=\frac{\sqrt{\beta_0}}{d_{i,m}}
\sqrt{\frac{K_R}{K_R+1}}e^{-j\frac{2\pi}{\lambda_g}(x_m^{\mathrm{PA}}-l_{m-1})}e^{-j\frac{2\pi}{\lambda}d_{i,m}},
\end{align}
where $d_{i,m}$ denotes the distance between client $i$ and the PA on segment $m$.

To characterize the joint effect of all segments, we collect the PA positions into
$\xb^{\mathrm{PA}}\triangleq[x_1^{\mathrm{PA}},\dots,x_M^{\mathrm{PA}}]^T$, and define the downlink channel vector and its LoS mean vector as
\begin{align}
&\hb_i(\xb^{\mathrm{PA}})=\left[h_{i,1}(x_1^{\mathrm{PA}}),
\dots,h_{i,M}(x_M^{\mathrm{PA}})\right]^T\in \mathbb{C}^{M\times 1}, \\
&\bar{\hb}_i(\xb^{\mathrm{PA}})=\left[\bar{h}_{i,1}(x_1^{\mathrm{PA}}),
\dots,\bar{h}_{i,M}(x_M^{\mathrm{PA}})\right]^T\in \mathbb{C}^{M\times 1}.
\end{align}
Under independent NLoS scattering across segments, let
\begin{align}
\Db_i(\xb^{\mathrm{PA}})\triangleq \diag\left(d_{i,1}^{-2},\ldots,d_{i,M}^{-2}\right).
\end{align}
Then, the Rician statistical channel correlation matrix is
\begin{align}
\Ab_i(\xb^{\mathrm{PA}})
&\triangleq \mathbb E\left[\hb_i^*(\xb^{\mathrm{PA}})\hb_i^T(\xb^{\mathrm{PA}})\right]  =\bar{\hb}_i^*(\xb^{\mathrm{PA}})\bar{\hb}_i^T(\xb^{\mathrm{PA}})
+\frac{\beta_0}{K_R+1}\Db_i(\xb^{\mathrm{PA}}).
\label{eq:rician_statistical_covariance}
\end{align}

Let $\wb^{\rm I}=[w_1^{\rm I},\dots,w_M^{\rm I}]^T\in\mathbb{C}^{M\times 1}$ and
$\xb^{{\rm PA},{\rm I}}=[x_1^{{\rm PA},{\rm I}},\dots,x_M^{{\rm PA},{\rm I}}]^T$ denote the beamforming vector and PA position vector used in the global-model multicast stage, respectively. Similarly, let
$\wb_i^{\rm II}=[w_{i,1}^{\rm II},\dots,w_{i,M}^{\rm II}]^T\in\mathbb{C}^{M\times 1}$ and
$\xb_i^{{\rm PA},{\rm II}}=[x_{i,1}^{{\rm PA},{\rm II}},\dots,x_{i,M}^{{\rm PA},{\rm II}}]^T$ denote the corresponding vectors used in the TDMA slot assigned to client $i$ in the control-variable unicast stage. 

If the server transmits normalized information-bearing symbols $s^{\rm I}$ and $s_i^{\rm II}$, with $\mathbb E[|s^{\rm I}|^2]=\mathbb E[|s_i^{\rm II}|^2]=1$, the received signals at client $i$ in the two stages are
\begin{equation}
\begin{aligned}
y_i^{D,{\rm I}}&=\hb_i^T(\xb^{{\rm PA},{\rm I}})\wb^{\rm I}s^{\rm I}+v_i^{\rm I},\\
y_i^{D,{\rm II}}&=\hb_i^T(\xb_i^{{\rm PA},{\rm II}})\wb_i^{\rm II}s_i^{\rm II}+v_i^{\rm II},
\end{aligned}
\end{equation}
where $v_i^{\rm I}$ and $v_i^{\rm II}$ are additive Gaussian noises distributed as $\mathcal{CN}(0,\sigma_D^2)$, and $\sigma_D^2=N_0W_D$ denotes the downlink noise power. Accordingly, for either stage-specific pair $(\wb,\xb^{\rm PA})$, the average received SNR of client $i$ is
\begin{align}
\bar{\nu}_i(\wb,\xb^{\mathrm{PA}})=\frac{\mathbb E\left[|\hb_i^T(\xb^{\mathrm{PA}})\wb|^2\right]}{\sigma_D^2}
=\frac{\wb^H\Ab_i(\xb^{\mathrm{PA}})\wb}{\sigma_D^2}.
\label{eq:average_downlink_snr}
\end{align}
The corresponding average-SNR-based downlink rate metric is defined as
\begin{align}
R_i^D(\wb,\xb^{\rm PA})=W_D\log_2\left(1+\bar{\nu}_i(\wb,\xb^{\mathrm{PA}})\right).
\label{eq:generic_downlink_rate}
\end{align}

\subsection{Downlink Problem Formulation}

The convergence of the global model in Algorithm \ref{alg: FedAVR} can be affected by quantization distortion and transmission failures. Accordingly, in each FL round, the objective of downlink transmission is to maximize the common number of downlink quantization bits  that can be supported by all active clients under the adopted average-SNR-based downlink rate metric and the prescribed downlink latency budget $\tau_D$.

Under the two-stage downlink protocol, the downlink latency budget $\tau_D$ is allocated across the two downlink transmission stages. Let $\tau^{\rm I}$ denote the duration of the global-model multicast stage and $\tau_i^{\rm II}$ denote the unicast duration assigned to client $i$ in the control-variable unicast stage. The two-stage downlink time allocation must satisfy
\begin{align} \label{eqn: time allocation}
    \tau^{\rm I}+\sum_{i\in\Sc^r}\tau_i^{\rm II} \leq \tau_D,~
    \tau^{\rm I}>0,~ \tau_i^{\rm II}>0,\ \forall i\in\Sc^r.
\end{align}
Accordingly, the downlink design seeks to maximize the downlink quantization bits $B_{D}$ via joint optimization of the two-stage transmission and time allocation, while supporting the delivery of both quantized $\thetab^r$ and its associated $\cb_i^r$ within the downlink latency budget $\tau_D$ under the average-SNR-based downlink rate metric. In the unicast stage, the server also sends a per-client control-signaling payload $b_{\rm ctl}$ to notify the required uplink transmission configuration, such as the assigned uplink resource and quantization bits.

Collect the downlink optimization variables for round $r$ as
\begin{align}
\mathcal{V}_D \triangleq\Big\{&B_D,\tau^{\rm I},
\wb^{\rm I},\xb^{{\rm PA},{\rm I}}, \{\tau_i^{\rm II},\wb_i^{\rm II},\xb_i^{{\rm PA},{\rm II}}\}_{i\in\Sc^r}
\Big\}.
\label{eq:downlink_variable_set}
\end{align}
Then, the downlink design problem is formulated as
\begin{subequations}\label{prob:two_phase_downlink}
\begin{align}
& \max_{\mathcal{V}_D} \ B_D \label{prob:dl_obj} \\
\text{s.t.} ~& d(B_D+1)+\mu\leq\tau^{\rm I} R_i^D(\wb^{\rm I},\xb^{{\rm PA},{\rm I}}), \forall i\in\Sc^r,\label{prob:dl_c1}\\
& d(B_D+1)+\mu+b_{\rm ctl}\leq\tau_i^{\rm II}  R_i^D(\wb_i^{\rm II},\xb_i^{{\rm PA},{\rm II}}),\forall i\in\Sc^r, \label{prob:dl_c2}\\
& \tau^{\rm I}+\sum_{i\in\Sc^r}\tau_i^{\rm II} \leq \tau_D,\tau^{\rm I}>0, \tau_i^{\rm II}>0, \forall i\in\Sc^r,\label{prob:dl_c3}\\
& \|\wb^{\rm I}\|^2 \leq P_{\max}, ~ \|\wb_i^{\rm II}\|^2 \leq P_{\max},~ \forall i\in\Sc^r, \label{prob:dl_c4}\\
& x_m^{{\rm PA},{\rm I}}\in\Lc_m, x_{i,m}^{{\rm PA},{\rm II}}\in\Lc_m,~\forall i\in\Sc^r, \forall m\in\Mc, \label{prob:dl_c5}\\
& B_D\in\Zbb,~ B_D\geq B_D^{\min}, \label{prob:dl_c6}
\end{align}
\end{subequations}
where $B_D^{\min}\geq 1$ is an integer denoting the minimum admissible downlink quantization precision required by FL, and $b_{\rm ctl}$ denotes the number of bits used to represent $\{W_i^r, B_i^r\}$. Here, constraints \eqref{prob:dl_c1} and \eqref{prob:dl_c2} respectively enforce the latency requirements for multicast and unicast delivery under the average-SNR-based downlink rate metric. Constraints \eqref{prob:dl_c4} impose the server transmit power budget $P_{\max}$ for the two stages, while constraints \eqref{prob:dl_c5} and \eqref{prob:dl_c6} specify the feasible range of each PA position and the quantization bits to be optimized $B_D$.

Solving problem \eqref{prob:two_phase_downlink} is nontrivial due to the tight coupling among integer quantization, time allocation, beamforming, and PA positioning. In particular, the integer constraint \eqref{prob:dl_c6} makes the problem mixed-integer, while the time allocation variables are coupled through the total latency budget and jointly determine the supportable quantization precision together with the rate metrics. Moreover, the rate metrics are highly non-convex in the beamforming vectors and PA positions, since the received signal is formed by coherent superposition of signals from multiple segments.

\vspace{-0.3cm}
\subsection{Rate-Oriented Solution Framework}

To address these coupled challenges, we develop a rate-oriented solution framework that exploits the separable structure of the two-stage downlink transmission. For a given design, let $R_{\rm mc}\triangleq\min_{i\in\Sc^r}R_i^D(\wb^{\rm I},\xb^{{\rm PA},{\rm I}})$ and $R_{{\rm uc},i}\triangleq R_i^D(\wb_i^{\rm II},\xb_i^{{\rm PA},{\rm II}})$ denote the multicast and client-$i$ unicast rate metrics, respectively. The framework is based on the following proposition.

\vspace{-0.2cm}
\begin{Proposition} \label{prop:downlink_time_bit_recovery} For any given downlink physical-layer design with positive multicast rate metric $ R_{\rm mc}$ and positive unicast rate metrics $\{R_{{\rm uc},i}\}_{i\in\Sc^r}$, the minimum stage durations required to support a fixed admissible downlink quantization bit $B_D$ are
\begin{align} 
&\tau^{{\rm I}\star}(B_{D}) = \frac{d(B_D+1)+\mu}{ R_{\rm mc}},\label{eq:downlink_tauI_star}  \\
&\tau_i^{{\rm II}\star}(B_D) = \frac{d(B_D+1)+\mu+b_{\rm ctl}}{ R_{{\rm uc},i}}, \forall i\in\Sc^r. \label{eq:downlink_tauII_star} 
\end{align} 
The integer upper bound on the supportable downlink quantization bit number under the latency budget $\tau_D$ is
\begin{align} 
B_D^{\star} = \left\lfloor \frac{1}{d} \left( \frac{\tau_D-b_{\rm ctl}\sum_{i\in\Sc^r}\frac{1}{R_{{\rm uc},i}}} { \frac{1}{ R_{\rm mc}} + \sum_{i\in\Sc^r}\frac{1}{R_{{\rm uc},i}} } -\mu \right) -1 \right\rfloor . \label{eq:downlink_bit_recovery} 
\end{align} 
For this design, if $B_D^{\star}\geq B_D^{\min}$, it is the maximum admissible bit number and the above durations attain it; otherwise, the design cannot support the minimum precision.
Moreover, if the multicast and unicast rate-maximization subproblems are globally solved, i.e.,
\begin{subequations}\label{prob:multicast_rate_opt}
\begin{align}
   R_{\rm mc}^\star
&=
\max_{\wb^{\rm I},\,\xb^{{\rm PA},{\rm I}}}
\min_{i\in\Sc^r}
R_i^D(\wb^{\rm I},\xb^{{\rm PA},{\rm I}})
\\
\text{\rm s.t.}~
& \|\wb^{\rm I}\|^2\le P_{\max}, x_m^{{\rm PA},{\rm I}}\in\Lc_m, \forall m\in\Mc, \label{prob:multicast_rate}
\end{align}
\end{subequations} and
\begin{subequations}
\begin{align}
   R_{{\rm uc},i}^\star
&=
\max_{\wb_i^{\rm II},\,\xb_i^{{\rm PA},{\rm II}}}
 R_i^D(\wb_i^{\rm II},\xb_i^{{\rm PA},{\rm II}})
\label{prob:unicast_rate_obj}\\
\text{\rm s.t.}~
& \|\wb_i^{\rm II}\|^2\le P_{\max}, x_{i,m}^{{\rm PA},{\rm II}}\in\Lc_m,\forall m\in\Mc, \label{prob:unicast_rate}
\end{align}
\end{subequations} 
then, if the resulting $B_D^{\star}\geq B_D^{\min}$, this physical-layer design together with \eqref{eq:downlink_tauI_star}, \eqref{eq:downlink_tauII_star} and \eqref{eq:downlink_bit_recovery} is globally optimal for problem \eqref{prob:two_phase_downlink}. Otherwise, problem \eqref{prob:two_phase_downlink} is infeasible.
\end{Proposition}

\noindent\emph{Proof:} See Appendix~\ref{app:proof_prop_downlink_time_bit_recovery}.

Proposition \ref{prop:downlink_time_bit_recovery} shows that problem \eqref{prob:two_phase_downlink} admits a rate-oriented solution structure. In particular, the original mixed-integer downlink design reduces to physical-layer rate optimization in the global-model multicast and control-variable unicast stages, after which the optimal time allocation and the maximum supportable downlink quantization bit number can be recovered in closed form.

Algorithm~\ref{alg:two_phase_downlink} applies this framework. It alternates between feasible multicast beamforming and PA-position updates to obtain a multicast design and its actual rate $\widehat R_{\rm mc}$, and independently optimizes each unicast stage. The multicast procedure uses relaxation, randomization, and a finite position grid, and is not guaranteed to attain the global rate maximum required by the proposition. The resulting rates are used to recover the bit number and durations as described in Secs.~\ref{sec: multicast design} and \ref{sec: unicast design}.

\begin{algorithm}[h!]
\caption{Two-Stage Downlink Transmission Design}
\label{alg:two_phase_downlink}
\begin{algorithmic}[1]
\STATE \textbf{Input:} $\Sc^r$, client coordinates $\{(x_i,y_i)\}$, segment search grids $\{\Xc_m\}$, $P_{\max}$, $W_D$, $\tau_D$, $B_D^{\min}$, and $b_{\rm ctl}$.

\STATE Initialize $\xb^{{\rm PA},{\rm I}}$ as the midpoint of its segment.
\REPEAT
    \STATE Solve the relaxed SDP in \eqref{prob:phase_I_sdr} and recover a feasible $\wb^{\rm I}$ by eigenvector decomposition or Gaussian randomization.
        \STATE Update $x_m^{{\rm PA},{\rm I}}$ by searching \eqref{eq:phase_I_position_update} over $\Xc_m$, $\forall m \in \Mc$.
\UNTIL{the stopping criterion is met}
\STATE Obtain a feasible $\wb^{{\rm I},r}$ and $\xb^{{\rm PA},{\rm I},r}$, and evaluate $\widehat R_{\rm mc}$ using \eqref{eq:multicast_rate_star}.

\FOR{each client $i\in\Sc^r$}
    \STATE Compute $\xb_i^{{\rm PA},{\rm II}\star}$, $\wb_i^{{\rm II}\star}$, $ R_{{\rm uc},i}^{\star}$ via \eqref{eq:phase_II_pa_closed_form}, principal-eigenvector beamforming, and \eqref{eq:phase_II_rate_star}.
\ENDFOR

\STATE Set $\wb_i^{{\rm II},r}=\wb_i^{{\rm II}\star},\xb_i^{{\rm PA},{\rm II},r}=\xb_i^{{\rm PA},{\rm II}\star}$.
\IF{$\widehat R_{\rm mc}\leq0$ or $R_{{\rm uc},i}^{\star}\leq0$ for some $i\in\Sc^r$}
    \STATE Report that the obtained downlink design is infeasible, skip this FL round, and terminate.
\ENDIF
\STATE Compute $B_D^r$ from \eqref{eq:downlink_bit_recovery} using $\widehat R_{\rm mc}$ and $\{R_{{\rm uc},i}^{\star}\}$.
\IF{$B_D^r<B_D^{\min}$}
    \STATE Report that the obtained downlink design is infeasible, skip this FL round, and terminate.
\ENDIF
\STATE Compute the minimum stage durations from \eqref{eq:downlink_tauI_star} and \eqref{eq:downlink_tauII_star}.

\STATE \textbf{Output:} $\wb^{{\rm I}, r}$, $\xb^{{\rm PA},{\rm I},r}$, $\{\wb_i^{{\rm II},r},\xb_i^{{\rm PA},{\rm II},r}\}_{i\in\Sc^r}$, and $B_D^r$.
\end{algorithmic}
\end{algorithm}

\vspace{-0.4cm}
\subsection{Multicast Beamforming and PA Positioning}
\label{sec: multicast design}

Since $R_i^D(\wb,\xb^{\rm PA})$ is monotonically increasing with the average received SNR by \eqref{eq:generic_downlink_rate}, the multicast rate-metric maximization problem in \eqref{prob:multicast_rate_opt} is equivalent to the following max-min average-SNR problem: 
\begin{subequations} \label{prob:phase_I_multicast} \begin{align} 
&\max_{\wb^{\rm I},\xb^{{\rm PA},{\rm I}},\nu^{\rm I}} ~ \nu^{\rm I} \label{prob:phase_I_obj}\\ \mathrm{s.t.}~ & (\wb^{\rm I})^H\Ab_i(\xb^{{\rm PA},{\rm I}})\wb^{\rm I} \geq \sigma_D^2\nu^{\rm I}, ~ \forall i\in\Sc^r, \label{prob:phase_I_c1}\\ 
& \|\wb^{\rm I}\|^2\leq P_{\max}, \label{prob:phase_I_c2}\\ & x_m^{{\rm PA},{\rm I}}\in\Lc_m, \forall m\in\Mc. \label{prob:phase_I_c3} 
\end{align} 
\end{subequations} 
where $\nu^{\rm I}$ denotes the minimum average received SNR in the global-model multicast stage. To obtain a computationally efficient solution, we decompose problem \eqref{prob:phase_I_multicast} into the beamforming and PA-positioning subproblems, which are solved via alternating optimization (AO) \cite{IterativeMMSE}.

\subsubsection{Beamforming Optimization for Given PA Positions}
For given PA position vector $\xb^{{\rm PA},{\rm I}}$, define $\Ab_i^{\rm I}\triangleq\Ab_i(\xb^{{\rm PA},{\rm I}})$ and $\Wb^{\rm I}\triangleq \wb^{\rm I}(\wb^{\rm I})^H$. The average received signal power of client $i$ admits an equivalent trace form, i.e., $(\wb^{\rm I})^H\Ab_i^{\rm I}\wb^{\rm I}=\trace(\Ab_i^{\rm I}\Wb^{\rm I})$.
Accordingly, the multicast beamforming optimization subproblem is given by
\begin{subequations}\label{prob:phase_I_sdr}
\begin{align}
& \max_{\Wb^{\rm I}\succeq 0,\nu^{\rm I}} \ \nu^{\rm I} \label{prob:phase_I_sdr_obj} \\
\text{s.t.} ~& \trace\left(\Ab_i^{\rm I}\Wb^{\rm I}\right)\geq
\nu^{\rm I}\sigma_D^2,~ \forall i\in\Sc^r,\label{prob:phase_I_sdr_c1}\\
& \trace(\Wb^{\rm I})\leq P_{\max}, \label{prob:phase_I_sdr_c2}\\
& \mathrm{Rank}(\Wb^{\rm I})=1. \label{prob:phase_I_sdr_c3}
\end{align}
\end{subequations}
The above subproblem is non-convex due to the rank-one constraint \eqref{prob:phase_I_sdr_c3}. We drop it and relax problem \eqref{prob:phase_I_sdr} into a standard semidefinite programming (SDP) \cite{SDRQOP_2010}, which can be efficiently solved via CVX. If the obtained solution is rank-one, we directly recover the beamforming vector by eigenvalue decomposition; otherwise, we apply Gaussian randomization to construct a feasible rank-one beamforming vector \cite{1634819}.

\subsubsection{PA Positioning for Given Beamforming Vector}
For given multicast beamforming vector $\wb^{\rm I}$, the PA positions are optimized in a segment-wise manner. When optimizing the position of the $m$-th PA, the other PA positions are fixed and denoted by $\xb_{-m}^{{\rm PA},{\rm I}}$. Accordingly, the optimization of $x_m^{\mathrm{PA}}$ is reduced to solving the max-min average-SNR subproblem:
\begin{align}
x_m^{{\rm PA},{\rm I}}=\arg\max_{x\in\Lc_m}\min_{i\in\Sc^r}
~(\wb^{\rm I})^H\Ab_i(x,\xb_{-m}^{{\rm PA},{\rm I}})\wb^{\rm I}.
\label{eq:phase_I_position_update}
\end{align}
Although subproblem \eqref{eq:phase_I_position_update} is non-convex, it involves only a scalar optimization variable. We approximate its solution by discretizing $\Lc_m$ into a uniform grid set $\Xc_m$ and performing exhaustive 1-D search over $\Xc_m$ \cite{MBO_pinching}.

The AO procedure alternates these updates until a stopping criterion is met. Since randomization and finite-grid positioning need not yield globally optimal updates, we evaluate the achievable multicast rate from the final feasible beamformer and PA positions:
\begin{align}
\widehat R_{\rm mc}=\min_{i\in\Sc^r}R_i^D(\wb^{{\rm I},r},\xb^{{\rm PA},{\rm I},r}).
\label{eq:multicast_rate_star}
\end{align}

\vspace{-0.6cm}
\subsection{Cooperative Unicast Design for Control-variable Delivery}
\label{sec: unicast design}

Since each unicast rate metric $ R_i^D(\wb_i^{\rm II},\xb_i^{{\rm PA},{\rm II}})$ is monotonically increasing with the average received SNR, the unicast rate-metric maximization for client $i$ is equivalent to
\begin{subequations}\label{prob:phase_II_unicast}
\begin{align}
& \max_{\wb_i^{\rm II},\xb_i^{{\rm PA},{\rm II}}} \
\frac{(\wb_i^{\rm II})^H\Ab_i(\xb_i^{{\rm PA},{\rm II}})\wb_i^{\rm II}}{\sigma_D^2}
\label{prob:phase_II_unicast_obj}\\
\text{s.t.} ~
& \|\wb_i^{\rm II}\|^2\leq P_{\max},~x_{i,m}^{{\rm PA},{\rm II}}\in\Lc_m, \forall m\in\Mc. \label{prob:phase_II_unicast_c2}
\end{align}
\end{subequations}

For any given PA position $\xb_i^{{\rm PA},{\rm II}}$, \eqref{prob:phase_II_unicast_obj} is a Rayleigh quotient, and the optimal
beamformer is the principal-eigenvector beamformer, i.e., $\wb_i^{{\rm II}\star} =\sqrt{P_{\max}}\vb_{\max}(\Ab_i)$, where $\vb_{\max}(\cdot)$ denotes the unit-norm eigenvector associated with
the largest eigenvalue $\lambda_{\max}(\Ab_i)$. Substituting this beamformer back into
\eqref{prob:phase_II_unicast_obj}, the maximum average received SNR becomes $P_{\max}\lambda_{\max}(\Ab_i)/\sigma_D^2$, which reduces \eqref{prob:phase_II_unicast} to the following PA-positioning
problem:
\begin{subequations}\label{prob:phase_II_pa_position}
\begin{align}
& \max_{\{x_{i,m}^{{\rm PA},{\rm II}}\}} \
\lambda_{\max}\left(\Ab_i(\xb_i^{{\rm PA},{\rm II}})\right)
\label{prob:phase_II_pa_obj}\\
\text{s.t.} ~
& x_{i,m}^{{\rm PA},{\rm II}}\in\Lc_m,~ \forall m\in\Mc. \label{prob:phase_II_pa_c1}
\end{align}
\end{subequations}

While \eqref{prob:phase_II_pa_obj} appears to couple the PA positions globally, it can in fact be decoupled. Specifically, by compensating for the deterministic phase shifts, $\Ab_i(\xb_i^{{\rm PA},{\rm II}})$ is unitarily similar to a real nonnegative matrix. Both its diagonal entries, corresponding to the segment-wise average channel power gains $\beta_0/d_{i,m}^2$, and its off-diagonal LoS correlation terms decrease as the segment-to-client distances ${d_{i,m}}$ increase. According to the Perron-Frobenius theorem, the largest eigenvalue $\lambda_{\max}(\mathbf{A}_i)$ is nondecreasing with respect to each entry of the equivalent nonnegative matrix. Therefore, simultaneously minimizing each distance $d_{i,m}$ yields a global maximizer of $\lambda_{\max}(\Ab_i(\xb_i^{{\rm PA},{\rm II}}))$.

Since the feasible regions $\Lc_m$ are mutually independent, the optimal PA position on segment $m$ is simply the geometric projection of client $i$'s coordinate $x_i$ onto $\Lc_m$:
\begin{align}
x_{i,m}^{{\rm PA},{\rm II}\star}=\Pi_{\Lc_m}(x_i)=\min\left\{\max\left\{x_i,l_{m-1}\right\},l_m\right\}.
\label{eq:phase_II_pa_closed_form}
\end{align}
Accordingly, the maximum unicast rate metric of client $i$ is 
\begin{align}
R_{{\rm uc},i}^{\star}=W_D\log_2\bigg(1+\frac{P_{\max}\lambda_{\max}(\Ab_i(\xb_i^{{\rm PA},{\rm II}\star}))}{\sigma_D^2}\bigg).
\label{eq:phase_II_rate_star}
\end{align}

\section{SWAN-ASSISTED UPLINK DESIGN}
\label{sec: uplink}

We now turn to the uplink transmission design. To facilitate FL, we adopt a parallel segment selection (PSS) protocol, where each active client is associated with only one serving segment. On this basis, we establish the uplink problem model and develop the  optimization frameworks for PA positioning, bandwidth allocation, and uplink quantization-bit design.

\vspace{-0.3cm}
\subsection{Uplink Signal Model}

Under the PSS protocol, each active client is pre-associated with exactly one serving segment according to its horizontal coordinate, and different segments can be activated simultaneously for parallel uplink transmission. Specifically, client $i$ is associated with segment $m$ if $x_i \in \Ic_m$. Accordingly, at round $r$, the set of active clients served by segment $m$ is defined as
\begin{align}
\Uc_m^r\triangleq\{i\in\Sc^r\mid x_i\in\Ic_m\}.
\label{eq:uplink_association}
\end{align}
Since each client is served by only one segment, no multi-segment coherent combining is involved in the uplink. Therefore, the uplink transmission design is determined by the link quality between each client and its serving PA.

For average-SNR-based uplink design, we characterize the uplink link quality through the average received SNR. Unlike the downlink cooperative beamforming design, the PSS protocol involves only a scalar client-segment link. Therefore, the Rician statistical effect reduces to the average channel power gain. Taking the expectation over the small-scale fading in \eqref{eq:general_channel}, the phase terms vanish after taking the squared magnitude, and the average channel power gain between client $i$ and the PA on segment $m$ is given by
\begin{align} \label{eqn: average_channel_gain}
    \mathbb{E}\big[|h_{i,m}(x_m^{\mathrm{PA}})|^2\big] = \frac{\beta_0}{(x_i - x_m^{\mathrm{PA}})^2 + y_i^2 + H^2}. 
\end{align}

Define $\rho_i(x) \triangleq (x_i - x)^2 + c_i, c_i \triangleq y_i^2 + H^2$, where $c_i$ captures the fixed lateral and vertical offset of client $i$ from the waveguide. Then, the carrier-to-noise-density ratio is 
\begin{align}
\gamma_{i,m}(x_m^{\mathrm{PA}})\triangleq\frac{P_i\beta_0}{N_0\rho_i(x_m^{\mathrm{PA}})},
\label{eq:sndr}
\end{align}
where $P_i$ denotes the transmit power of client $i$, and $N_0$ is the noise power spectral density.
Based on the above definitions, the average-SNR-based uplink transmission rate metric of client $i\in\mathcal{U}_m^r$ is defined as
\begin{align}
R_i(W_i^r,x_m^{\mathrm{PA}})=W_i^r\log_2\left(1+\frac{\gamma_{i,m}(x_m^{\mathrm{PA}})}{W_i^r}\right),
\label{eq:rate_general}
\end{align}
where $W_i^r$ is the bandwidth allocated to client $i$ in round $r$.

\vspace{-0.4cm}
\subsection{Uplink Problem Formulation}

In FL, global aggregation can be performed only after all active clients complete the upload of their local updates within the current communication round. Consequently, the round latency is determined by the client with the weakest uplink transmission capability. Under a common uplink latency budget, this straggler also dictates the minimum quantization level that can be supported in the system. This motivates a fairness-oriented uplink design that accounts for the worst-case client.

To account for worst-case uplink clients, we adopt a max-min fairness criterion, which optimizes the bandwidth allocation $\{W_i\}$, PA positions $\{x_m^{\mathrm{PA}}\}$, and uplink quantization bits $\{B_i\}$, so as to maximize the minimum quantization level among all active clients under the uplink latency budget $\tau_U$. Formally, the optimization problem is formulated as

\vspace{-0.6cm}
\begin{subequations}\label{prob:p_general}
\begin{align}
& \max_{\{W_i\},\, \{x_m^{\mathrm{PA}}\},\, \{B_i\}}\min_{i\in\Sc^r} B_i
\label{prob:p_obj} \\
\text{s.t.} ~
& \frac{d(B_i+1)+\mu}{R_i(W_i,x_m^{\mathrm{PA}})}\leq \tau_U,~
 \forall i\in\Uc_m^r, \forall m\in\Mc,
\label{prob:p_c1}\\
& \sum_{m \in \Mc} \sum_{i\in\Uc_m^r}W_i\leq W_{U}, W_i>0, \forall i\in\Sc^r,
\label{prob:p_c2}\\
& x_m^{\mathrm{PA}}\in\Lc_m,~ \forall m\in\Mc,
\label{prob:p_c3}\\
& B_i\in\mathbb{Z},~ B_i\geq B_U^{\min},~ \forall i\in\Sc^r.
\label{prob:p_c4}
\end{align}
\end{subequations}
\vspace{-0.6cm}

\noindent Here, constraint \eqref{prob:p_c1} imposes the latency requirement for uploading the quantized local update of each active client. Constraint \eqref{prob:p_c2} enforces the total bandwidth budget $W_U$ and enforces positive bandwidth allocation for all active clients. Constraint \eqref{prob:p_c3} specifies the feasible movement region of each PA, while constraint \eqref{prob:p_c4} requires the uplink quantization bits to be integer-valued and no smaller than the minimum admissible integer precision $B_U^{\min}\geq1$ for FL aggregation.

Problem \eqref{prob:p_general} is challenging to solve due to its max-min, mixed-integer and non-convex nature. To obtain tractable solutions, we progressively investigate three cases of increasing complexity: 1) PA positioning with fixed bandwidth and quantization; 2) joint PA positioning and heterogeneous quantization optimization; 3) joint bandwidth allocation, PA positioning, and quantization optimization.

\vspace{-0.2cm}
\subsection{PA Positioning with Fixed Bandwidth and Quantization}\label{sec:case1}

We start with a simplified setup where the total bandwidth is equally allocated among the active clients, i.e.,
$W_i^r = W_{\mathrm{eq}}= W_{U}/|\Sc^r|$, and the uplink quantization precision is fixed to a prescribed integer value $B_{\mathrm{fix}}$, i.e., $B_i^r =B_{\mathrm{fix}} \geq B_U^{\min} $, $\forall i\in\Sc^r$. Under this equal-bandwidth and fixed-quantization policy, the coupling between radio resource allocation and PA positioning is removed, allowing us to focus solely on the geometric optimization of the PA positions.

For fixed $B_{\mathrm{fix}}$ and $W_{\mathrm{eq}}$, problem \eqref{prob:p_general} reduces to a feasibility subproblem, namely checking whether the latency constraint \eqref{prob:p_c1}  can be satisfied by proper PA positioning. For each segment $m$, it amounts to maximizing the worst-case rate margin among the associated clients. Specifically, define the rate margin of client $i$ as the difference between the achievable uplink rate and the target one induced by the fixed quantization level $B_{\mathrm{fix}}$, i.e.,
\begin{align}
\phi_i(x)\triangleq W_{\mathrm{eq}} \log_2\left(1+\frac{\Gamma_i}{\rho_i(x)}\right) -\frac{d(B_{\mathrm{fix}}+1)+\mu}{\tau_U},
\label{eqn: max_rate_margin}
\end{align}
where $\Gamma_i \triangleq \frac{P_i\beta_0}{N_0W_{\mathrm{eq}}}$. Moreover, since $\phi_i(x)$ is monotonically decreasing with respect to $\frac{\rho_i(x)}{\Gamma_i}$, the PA-positioning problem for segment $m$ is given by
\begin{align}
\min_{x\in\Lc_m}\max_{i\in\mathcal{U}_m^r}\frac{\rho_i(x)}{\Gamma_i}.
\label{eq:case1_equiv}
\end{align}

Intriguingly, this reformulation \eqref{eq:case1_equiv} admits an exact closed-form finite-candidate solution without spatial discretization, as stated in the following theorem. 

\vspace{-0.3cm}
\begin{Theorem}
\label{thm:2d_pa}
For segment $m$ with $\mathcal{U}_m^r\neq\emptyset$, define $\tilde g_i(x)\triangleq\rho_i(x)/\Gamma_i$ and retain one index for each distinct function in the index set $\widetilde{\Uc}_m^r\subseteq\Uc_m^r$. Every globally optimal PA position of problem \eqref{eq:case1_equiv} belongs to the finite candidate set
\begin{align}
\mathcal{C}_m\triangleq\mathcal{C}_m^{\mathrm{pair}}\cup\mathcal{C}_m^{\mathrm{apex}}\cup\mathcal{C}_m^{\mathrm{bdry}},
\end{align}
where the three subsets are defined as follows:
\begin{enumerate}[\upshape (i)]
\item \emph{Pairwise intersections}: 
\begin{align}
\Cc_m^{\mathrm{pair}}
=
\{
x\in\Rbb\ | &\exists i,j\in\widetilde{\Uc}_m^r,i<j,
(\Gamma_j-\Gamma_i)x^2\notag \\
&-2(\Gamma_jx_i-\Gamma_ix_j)x+\Gamma_j(x_i^2+c_i)-\Gamma_i(x_j^2+c_j)=0
\}
\end{align}

\item \emph{Apex points}: $\Cc_m^{\mathrm{apex}}=\{x_i\mid i\in\mathcal{U}_m^r\}$;
\item \emph{Boundary points}: $\Cc_m^{\mathrm{bdry}}=\{l_{m-1},l_m\}$.
\end{enumerate}
The optimal PA position is then obtained by
\begin{align}
x_m^{\mathrm{PA} \star}=\arg\min_{x\in\mathcal{C}_m\cap\Lc_m}\max_{i\in\mathcal{U}_m^r}\frac{\rho_i(x)}{\Gamma_i}.
\label{eq:pa_opt}
\end{align}
\end{Theorem}

\noindent\emph{Proof:} See Appendix~\ref{app:proof_thm1}.

Theorem \ref{thm:2d_pa} shows that the optimal PA positions $\{x_m^{\rm PA \star}\}$ to problem \eqref{eq:case1_equiv} can be found exactly by evaluating candidate points in $\Cc_{m}$, and picking the one that minimizes the upper envelope in \eqref{eq:pa_opt}, without resorting to iterative optimization or spatial discretization for grid search. 

The prescribed equal-bandwidth and fixed quantization setting is feasible for segment $m$ if and only if the resulting worst-case rate margin is nonnegative with $x_m^{\rm PA\star}$, i.e., $\min_{i\in\mathcal{U}_m^r}\phi_i(x_m^{\mathrm{PA}\star})\geq 0$.

\begin{algorithm}[h!]
\caption{PA Positioning and Heterogeneous Quantization}
\label{alg:case2}
\begin{algorithmic}[1]
\STATE \textbf{Input:} the set of active clients $\Sc^r$, client coordinates $\{(x_i,y_i)\}$, transmit powers $\{P_i\}$, $W_U$, $\tau_U$, and $B_U^{\min}$.
\STATE Set $W_i^r=W_{\mathrm{eq}}=W_U/|\Sc^r|$, $\forall i\in\Sc^r$.
\FOR{$m=1$ to $M$}
  \IF{$\mathcal{U}_m^r=\emptyset$}
    \STATE Choose any $x_m^{\mathrm{PA},r}\in\Lc_m$.
  \ELSE
    \STATE Construct the candidate set $\mathcal{C}_m$ via Theorem~\ref{thm:2d_pa}.
    \STATE Obtain $x_m^{\mathrm{PA}\star}$ by solving \eqref{eq:pa_opt} with $\mathcal{C}_m$.
    \STATE Set $x_m^{\mathrm{PA},r}=x_m^{\mathrm{PA}\star}$.
    \FOR{each $i\in\mathcal{U}_m^r$}
        \STATE Set $B_i^r=\lfloor \bar B_i(x_m^{\mathrm{PA}\star})\rfloor$ via \eqref{eq:continuous_bit_case2}.
    \ENDFOR
  \ENDIF
\ENDFOR
\IF{$B_i^r<B_U^{\min}$ for some $i\in\Sc^r$}
    \STATE Report equal-bandwidth design infeasibility, skip this FL round, and terminate.
\ENDIF
\STATE \textbf{Output:} $\{x_m^{\mathrm{PA},r}\}$, $\{W_i^r\}$, and $\{B_i^r\}$ for FL round $r$.
\end{algorithmic}
\end{algorithm}
\vspace{-0.3cm}

\vspace{-0.4cm}
\subsection{Joint PA Positioning and Quantization Optimization}
\label{sec:case2}

The fixed-quantization design above is dictated by the worst-case client in each segment. Although this uniform policy simplifies the uplink design, it may leave substantial channel capacity of non-straggler clients underutilized. In FL, this is undesirable because clients with better channels could otherwise transmit higher-precision updates and thus reduce aggregation distortion. Motivated by this, we now consider solving problem \eqref{prob:p_general} through joint optimization of the PA positions and the client-adaptive uplink quantization bits under the equal-bandwidth policy.

We first relax the integer constraints \eqref{prob:p_c4} and consider their continuous counterparts. For a given PA position $x$, each client's precision can be raised to its individual maximum without reducing the max-min objective. This choice makes its latency constraint tight, although other optimal allocations may leave slack for non-straggler clients. The maximum precision supported by client $i\in\mathcal{U}_m^r$ at position $x$ is
\begin{align}
 \bar B_i(x) =A\log_2\left(1+\frac{\Gamma_i}{\rho_i(x)}\right)-C_0,
\label{eq:continuous_bit_case2}
\end{align}
where $A = \tau_U W_{\mathrm{eq}}/d$ and $C_0 = 1 + \mu/d$. 

Since $A$ and $C_0$ are identical for all clients, $\bar B_i(x)$ is strictly increasing with respect to $\Gamma_i/\rho_i(x)$. Therefore, for each segment $m$, maximizing the minimum supportable continuous quantization precision in problem \eqref{prob:p_general} is equivalent to
\begin{align}
\min_{x\in\Lc_m}\max_{i\in\mathcal{U}_m^r}\frac{\rho_i(x)}{\Gamma_i},
\label{eq:case2_equiv}
\end{align}
which is exactly the same PA positioning problem to \eqref{eq:case1_equiv} under fixed bandwidth and fixed quantization. Hence, the optimal PA position $\{x_m^{\mathrm{PA}\star}\}$ can be obtained from the closed-form characterization in Theorem~\ref{thm:2d_pa}.

Once $x_m^{\mathrm{PA}\star}$ is determined, the supportable continuous quantization precision of client $i\in\mathcal{U}_m^r$ is given by $\bar B_i(x_m^{\mathrm{PA}\star})$.
The corresponding integer-valued uplink quantization bit number for round $r$ is recovered as $ B_i^r = \lfloor \bar B_i(x_m^{\mathrm{PA}\star})\rfloor$. 
The detailed procedure is summarized in Algorithm~\ref{alg:case2}.

\vspace{-0.2cm}
\subsection{Joint Bandwidth Allocation, PA Positioning, and Quantization Optimization}\label{sec:case3}

While the previous two designs assume equal-bandwidth allocation to simplify the uplink design, they can be overly conservative. In particular, allocating the same bandwidth to clients with highly different channel conditions may lead to inefficient use of the total bandwidth budget. To better exploit channel heterogeneity and further improve the minimum supportable uplink quantization level, we now turn to the general design that jointly optimizes the bandwidth allocation, PA positions, and quantization bits.

To handle the max-min objective in \eqref{prob:p_general}, we introduce a continuous auxiliary variable $\bar B_U$ to represent a common lower bound for the uplink quantization levels across all active clients, i.e., $\bar B_U \leq B_i, \forall i \in \Sc^r$ \cite{BK:BoydV04}. Since maximizing the worst-case quantization level is mathematically equivalent to maximizing $\bar B_U$, the integer max-min problem \eqref{prob:p_general} can be relaxed into the following continuous optimization problem:
\begin{subequations}\label{prob:p3}
\begin{align}
& \max_{\{W_i\},\,\{x_m^{\mathrm{PA}}\},\,\bar B_U} \ \bar B_U
\label{prob:p3_obj} \\
\text{s.t.} ~& \frac{d(\bar B_U+1)+\mu}
{R_i(W_i,x_m^{\mathrm{PA}})}\leq \tau_U,\forall i\in\mathcal{U}_m^r,\forall m\in\Mc,
\label{prob:p3_c1}\\
& \sum_{m=1}^{M}\sum_{i\in\mathcal{U}_m^r} W_i\leq W_{U}, ~ W_i>0,~\forall i\in\Sc^r,
\label{prob:p3_c2}\\
& x_m^{\mathrm{PA}}\in\Lc_m,~ \forall m\in\Mc,
\label{prob:p3_c3}\\
& \bar B_U\geq B_U^{\min} \geq 1.
\label{prob:p3_c5}
\end{align}
\end{subequations}
Note that, unlike the equal-bandwidth cases, problem \eqref{prob:p3} involves a globally coupled bandwidth allocation across all segments, as seen in \eqref{prob:p3_c2}. In addition, the corresponding PA-positioning subproblem no longer admits the segment-wise max-min structure. Hence, the closed-form solution developed in Theorem~\ref{thm:2d_pa} cannot be directly applied.

To handle this global coupling, we rewrite the constraints \eqref{prob:p3_c1} and \eqref{prob:p3_c2} from an inverse perspective. Specifically, we fix a target $\bar B_U$ and characterize the minimum total bandwidth required to support it. To this end, define
\begin{align}
S(\bar B_U)&\triangleq\frac{d(\bar B_U+1)+\mu}{\tau_U},
\label{eq:case3_required_rate}\\
C(W,\gamma)&\triangleq W\log_2\left(1+\gamma/W\right),
\label{eq:St}
\end{align}
where $S(\bar B_U)$ denotes the minimum uplink transmission rate required by the quantization level $\bar B_U$, and $C(W,\gamma)$ denotes the uplink transmission rate under bandwidth $W$ and the channel-quality parameter $\gamma$\footnote{For any $\gamma>0$, $C(W,\gamma)$ is strictly increasing in $W$, with $\lim_{W\to\infty}C(W,\gamma)=\gamma/\ln 2$. Hence, its inverse function with respect to $W$, denoted by $C^{-1}(S,\gamma)$, is well defined over $0<S<\gamma/\ln 2$. If $S\geq\gamma/\ln 2$, we set $C^{-1}(S,\gamma)=+\infty$.}. Since $C(W,\gamma)$ is strictly increasing in $W$, its inverse $C^{-1}(S,\gamma)$ gives the minimum bandwidth required to support rate $S$ under channel condition $\gamma$. Therefore, for any fixed PA deployment, the minimum bandwidth required by client $i \in \Uc_{m}^r$ to support $\bar B_U$ is 
\begin{align}
    C^{-1} \left( S(\bar B_U), \gamma_{i,m}(x_m^{\mathrm{PA}}) \right),
\end{align}
and the minimum total bandwidth required to support $\bar B_U$ is given by minimizing the sum of these per-client minimum bandwidths over all feasible PA positions, i.e.,
\begin{align}
    G(\bar B_U) =& \min_{\substack{x_m^{\mathrm{PA}}\in\Lc_m,\forall m\in\Mc}}\sum_{m=1}^{M}   \sum_{i\in\Uc_m^r} C^{-1} ( S(\bar B_U), \gamma_{i,m}(x_m^{\mathrm{PA}})) \notag \\
    =& \sum_{m=1}^{M} \min_{x_m^{\mathrm{PA}}\in \Lc_m}  \sum_{i\in\Uc_m^r} C^{-1} ( S(\bar B_U), \gamma_{i,m}(x_m^{\mathrm{PA}})), \label{eqn: min_bandwidth}
\end{align}
where \eqref{eqn: min_bandwidth} follows because each subproblem depends only on the PA position and clients of segment $m$.

The following proposition provides an equivalent reformulation of problem \eqref{prob:p3} by replacing its coupled feasibility constraints with a single minimum-bandwidth condition, which also admits an explicit optimal characterization.

\vspace{-0.2cm}
\begin{Proposition} \label{prop:case3_inverse_bandwidth} 
For any prescribed $\bar B_U\geq B_U^{\min}$, the constraints \eqref{prob:p3_c1}--\eqref{prob:p3_c3} in problem \eqref{prob:p3} are feasible if and only if
\begin{align} 
G(\bar B_U)\leq W_U. \label{eq:case3_inverse_feasibility} 
\end{align} 
Consequently, problem \eqref{prob:p3} is equivalent to 
\begin{align} \label{prob:inverse_prob}
\bar B_U^{\star} = \max_{\bar B_U}\  \bar B_U, \quad {\rm s.t.}\quad G(\bar B_U)\leq W_U, \quad \bar B_U\geq B_U^{\min}.
\end{align} 
When the problem is feasible, one set of optimal PA positions and bandwidth allocation is
\begin{align} 
x_m^{\mathrm{PA}\star} &\in \mathop{\arg\min}_{x\in\Lc_m} \sum_{i\in\mathcal{U}_m^r} C^{-1} \big( S(\bar B_U^{\star}), \gamma_{i,m}(x) \big), \label{eq:case3_opt_pa}\\ 
W_i^{\star} &= C^{-1} \big( S(\bar B_U^{\star}), \gamma_{i,m}(x_m^{\mathrm{PA}\star}) \big), \forall m\in\Mc, i\in\mathcal{U}_m^r. \label{eq:case3_opt_bw} 
\end{align}

\end{Proposition} 

\noindent\emph{Proof:} See Appendix~\ref{app:proof_case3_inverse_bandwidth}.

Compared with problem \eqref{prob:p3}, problem \eqref{prob:inverse_prob} is much more tractable due to two-fold reasons. First, it eliminates the original coupling between bandwidth allocation and feasibility verification. Second, as $G(\bar B_U)$ is nondecreasing with respect to $\bar B_U$, the optimal quantization level $\bar B_U^{\star}$ can be efficiently obtained by bisection over $\bar B_U$.

For numerical implementation, Algorithm~\ref{alg:case3} uses finite nonempty position grids $\Xc_m\subseteq\Lc_m$. Define the grid-based required bandwidth by
\begin{align}
\widehat G(B)\triangleq\sum_{m=1}^{M}\min_{x\in\Xc_m}
\sum_{i\in\Uc_m^r} C^{-1}\!\left(S(B),\gamma_{i,m}(x)\right).
\label{eq:case3_grid_bandwidth}
\end{align}
The inner search evaluates $\widehat G$, with infeasible grid points assigned infinite cost; in general $\widehat G(B)\geq G(B)$. The outer layer applies bisection to the grid-based feasibility test.
To initialize the bisection search, we
construct a finite upper bound on any feasible value of $\bar B_U$.
Since $C(W,\gamma)<\gamma/\ln 2$ for any finite $W>0$, feasibility
requires
\begin{align*}
S(\bar B_U)<\min_{m:\mathcal{U}_m^r\neq\emptyset}
\min_{i\in\mathcal{U}_m^r}
\frac{\gamma_{i,m}^{\max}}{\ln 2},
\end{align*}
where $\gamma_{i,m}^{\max}=\max_{x\in\Lc_m}\gamma_{i,m}(x)$.
Substituting the definition of $S(\bar B_U)$ gives the bisection
interval $[B_U^{\min},B_{\mathrm{ub}}]$, with
\begin{align}
B_{\mathrm{ub}} = \frac{\tau_U}{d} \min_{m:\mathcal{U}_m^r\neq\emptyset} \min_{i\in\mathcal{U}_m^r} \frac{\gamma_{i,m}^{\max}}{\ln 2} -1-\frac{\mu}{d},
\label{eq:uplink_bit_upper_bound}
\end{align}
First test $\widehat G(B_U^{\min})\le W_U$. If it fails, the chosen grid cannot support the minimum precision, although a continuous-position solution may still exist. Otherwise, for each midpoint $B_{\mathrm{mid}}$, update the lower endpoint when $\widehat G(B_{\mathrm{mid}})\le W_U$, and the upper endpoint otherwise. Return the feasible lower endpoint $\widehat B_U$ at the prescribed tolerance. Recover each PA position from the corresponding grid minimizer in \eqref{eq:case3_grid_bandwidth}, and its client bandwidths from \eqref{eq:case3_opt_bw} evaluated at $\widehat B_U$. Finally, set $B_i^r=\lfloor\widehat B_U\rfloor$ for all $i\in\Sc^r$. This yields a feasible integer allocation; finite grid resolution and bisection tolerance need not yield the continuous or grid-based integer optimum.

\begin{algorithm}[h!]
\caption{Joint Bandwidth Allocation, PA Positioning and Quantization Optimization}
\label{alg:case3}
\begin{algorithmic}[1]
\STATE \textbf{Input:} $\Sc^r$, client coordinates $\{(x_i,y_i)\}$, transmit powers $\{P_i\}$, $W_U$, $\tau_U$, search grids $\{\Xc_m\}$, and $B_U^{\min}$.
\STATE Compute the upper bound $B_{\mathrm{ub}}$ according to \eqref{eq:uplink_bit_upper_bound}.
\IF{$B_{\mathrm{ub}}\leq B_U^{\min}$ or $\widehat G(B_U^{\min})>W_U$}
    \STATE Report grid-design infeasibility, skip this FL round, and terminate.
\ENDIF
\STATE Bisect $[B_U^{\min},B_{\mathrm{ub}}]$ using $\widehat G(B)\leq W_U$; return the feasible lower endpoint $\widehat B_U$.
\STATE Choose $\{x_m^{\mathrm{PA},r}\}$ as grid minimizers in \eqref{eq:case3_grid_bandwidth} at $\widehat B_U$.
\STATE Set $W_i^r=C^{-1}(S(\widehat B_U),\gamma_{i,m}(x_m^{\mathrm{PA},r}))$ and $B_i^r=\lfloor\widehat B_U\rfloor$ for $i\in\Uc_m^r$, $m\in\Mc$.
\STATE \textbf{Output:} $\{x_m^{\mathrm{PA},r}\}$, $\{W_i^r\}$, and $\{B_i^r\}$ for FL round $r$.
\end{algorithmic}
\end{algorithm}

\vspace{-0.2cm}
\section{Experiment Results} \label{sec: sim results}

\subsection{Experiment Setup}

We consider a wireless edge network over a $100$m $\times$ $80$m rectangular service area, where the BS and the SWAN waveguide are deployed at height $H=3$m. The base station is located at $(0,0,H)$ and equipped with a uniform linear array of $M_{\rm BS}=4$ antennas along the $x$-axis with half-wavelength spacing $d_{\rm BS}=\lambda/2$. The SWAN waveguide of length $L=100$m is deployed along the $x$-axis with its left end connected to the BS, and is uniformly partitioned into $M=4$ segments with boundaries [$0$m, $25$m, $50$m, $75$m, $100$m]. A total of $N=100$ clients are distributed over the service area according to a segment-wise clustered distribution, with one randomly generated cluster center in each SWAN segment. The carrier frequency is $f_c=28$ GHz, the relative dielectric constant of the waveguide is $\epsilon_r=1.96$, the reference channel gain is $\beta_0=\frac{c^2}{16f_c^2\pi^2}$ and  Rician factor is $K_R=10$ \cite{SWAN-2026,samy2025pass_ris_mmwave}.

For uplink and downlink transmissions, the total bandwidths are set to $W_U=W_D=100$ MHz, and the noise power spectral density is $N_0=2\times10^{-18}$ W/Hz. The transmit power of each client is uniformly drawn from [$0.1$ W, $0.5$ W], while the server transmit power budget is $P_{\max}=1$ W. The minimum admissible uplink and downlink quantization precisions are set to $B_U^{\min}=2$ and $B_D^{\min}=7$, respectively.

The image classification task on the widely used CIFAR-10 dataset is considered for evaluation. The adopted DNN model consists of four $3\times 3$ convolutional layers with channel sizes $64$, $128$, $256$, and $512$, each followed by $2\times 2$ max-pooling, and five fully connected layers with widths $128$, $256$, $512$, $1024$, and $10$. 
To simulate non-i.i.d. data, the CIFAR-10 training set is partitioned across $100$ clients following a Dirichlet distribution with hyperparameter $dir$ \cite{FedQVR_Shuai_2025}. Note that $dir$ controls the non-i.i.d. degree across clients, where a smaller value indicates a higher level of heterogeneity, and we set $dir=0.2$ by default, which implies a high level of data heterogeneity. The CIFAR-10 testing set is used to evaluate model performance after FL training.

We compare the proposed algorithm against three baselines, namely, FedAvg \cite{FedAvg_noniid_2019}, FedAdam \cite{AdaptiveFL_2021}, and SCAFFOLD \cite{karimireddy2020scaffold}, where the latter two are specifically designed to handle data heterogeneity \cite{FedQVR_Shuai_2025}. For the proposed algorithm, we further evaluate the three uplink designs in Sec. \ref{sec: uplink}. To ensure fairness, the same dual-quantization mechanism is applied to all baselines. In addition, three communication configurations are considered to examine the impact of different communication architectures and resource-allocation strategies on FL\footnote{For the baselines, the downlink adopts only multicast transmission, since they broadcast only the global model. By contrast, under \textbf{BS-OPT}, the proposed algorithm retains the same two-stage downlink transmission structure.}:
\begin{itemize}
    \item \textbf{FIX-QB}: an ideal-communication benchmark, where all active clients are assumed to complete both downlink and uplink transmissions successfully, and fixed downlink/uplink quantization bits are adopted without communication-aware resource optimization;
    \item \textbf{BS-OPT}: a conventional BS-based setting, where downlink/uplink transmissions are handled by the BS, and communication resources are optimized accordingly \cite{FedQVR_Shuai_2025}. It adopts the same total  bandwidth, latency budget, transmit powers, and joint bandwidth-allocation principle as the uplink and downlink designs in \textbf{SWAN-OPT}.
    \item \textbf{SWAN-OPT}: the SWAN-based setting, where both downlink and uplink transmissions are enabled by PAs. In this case, the SWAN-assisted two-stage downlink design and the uplink resource optimization developed in Secs.~IV and V are jointly employed.
\end{itemize}
  \begin{figure}[!t]
\centering
\subfigure[$B_i^r=2, B_D^r=6$]{\includegraphics[width=0.30\textwidth]{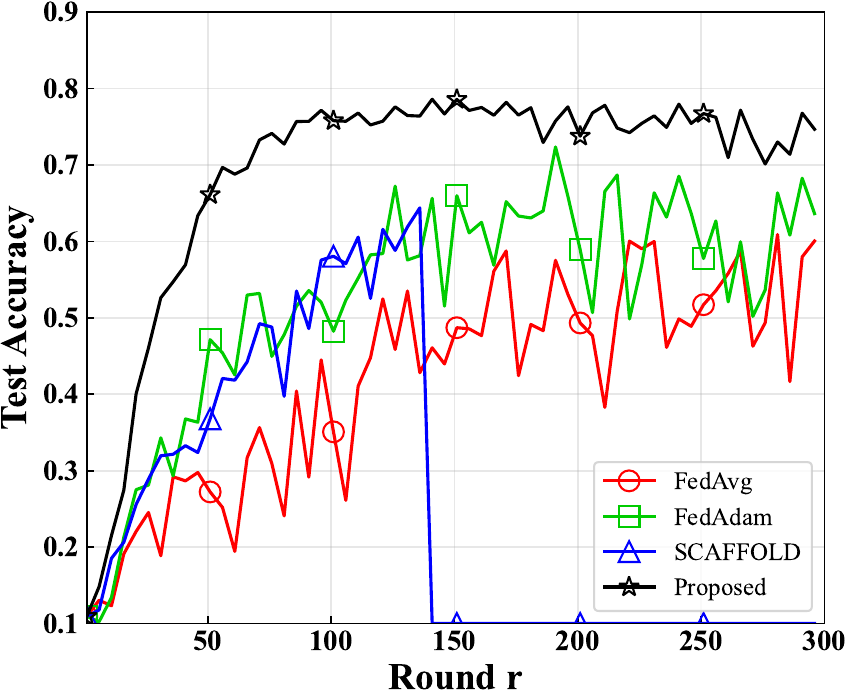}}
\subfigure[$B_i^r=2, B_D^r=7$]{\includegraphics[width=0.30\textwidth]{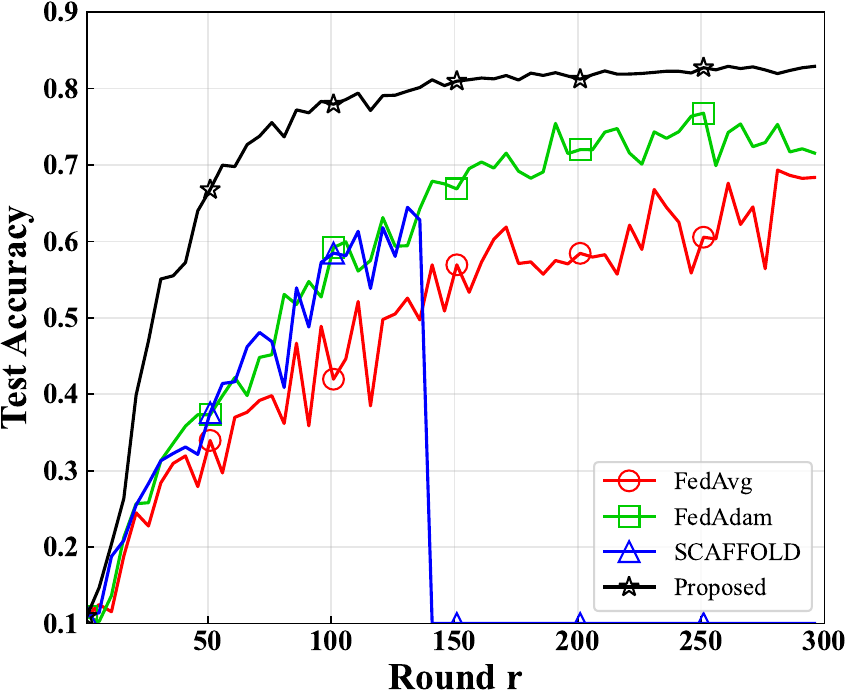}}
\subfigure[$B_i^r=2, B_D^r=8$]{\includegraphics[width=0.30\textwidth]{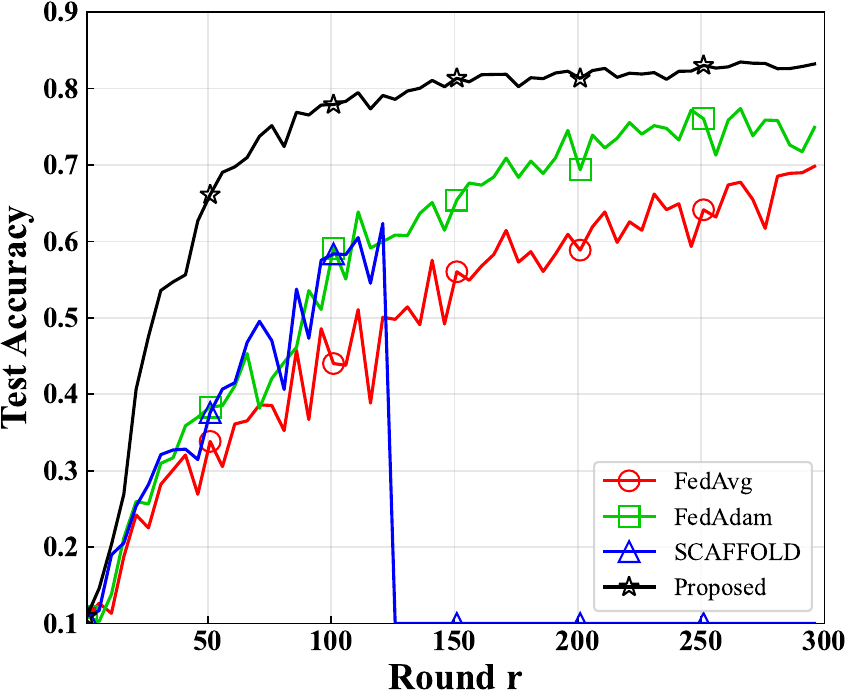}}\\
\vspace{-0.2cm}
\subfigure[$B_i^r=4, B_D^r=6$]{\includegraphics[width=0.30\textwidth]{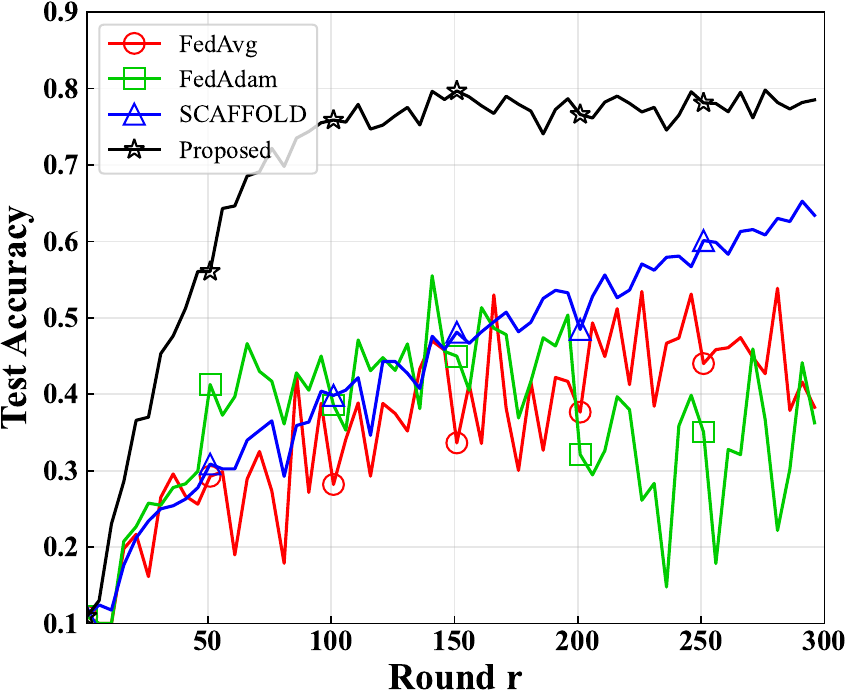}}
\subfigure[$B_i^r=4, B_D^r=7$]{\includegraphics[width=0.30\textwidth]{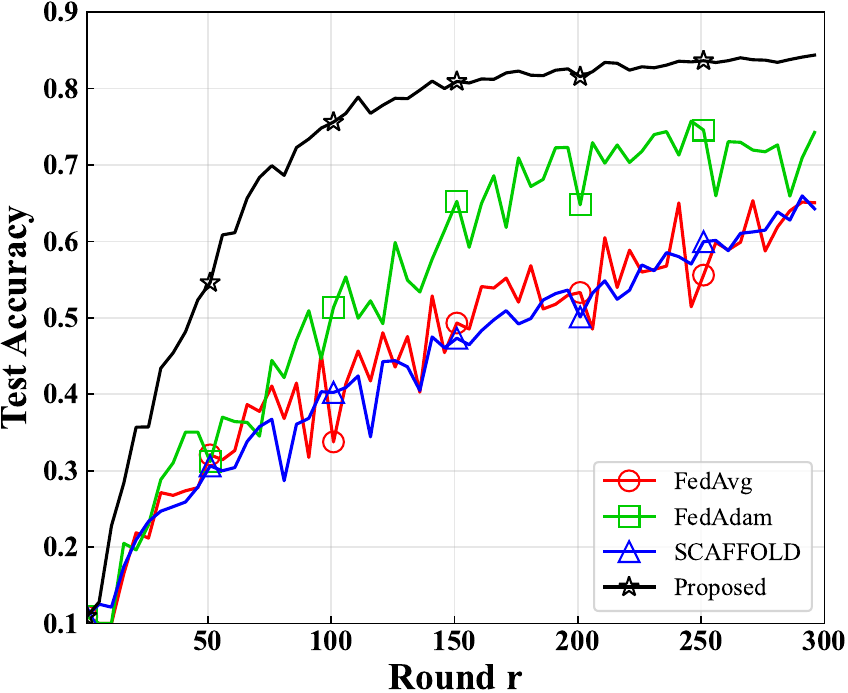}}
\subfigure[$B_i^r=4, B_D^r=8$]{\includegraphics[width=0.30\textwidth]{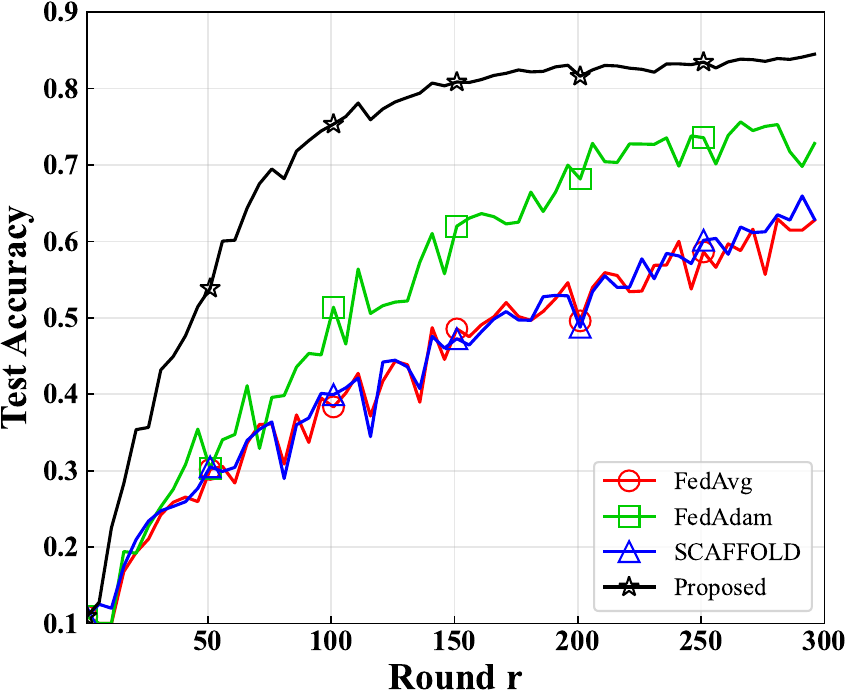}}
\vspace{-0.2cm}
\caption{Performance comparison under different quantization levels.}
\label{fig:quantization_comparison}
\end{figure}

 \textbf{Parameter setting}: For all algorithms with different communication configurations, $|\Sc^r|=10$ clients are sampled in each round, each of which performs local update of $3$ local epochs.  They employ a learning rate of $0.01$, a batch size of $50$.  The maximum number of communication rounds is $R= 300$. For the proposed method, the parameters $\gamma$ is $0.3$ and each local stepsize $a_i$ is $0.3$.

\begin{figure}[!th]
\centering
\subfigure[$dir=0.1$]{\includegraphics[width=0.44\textwidth]{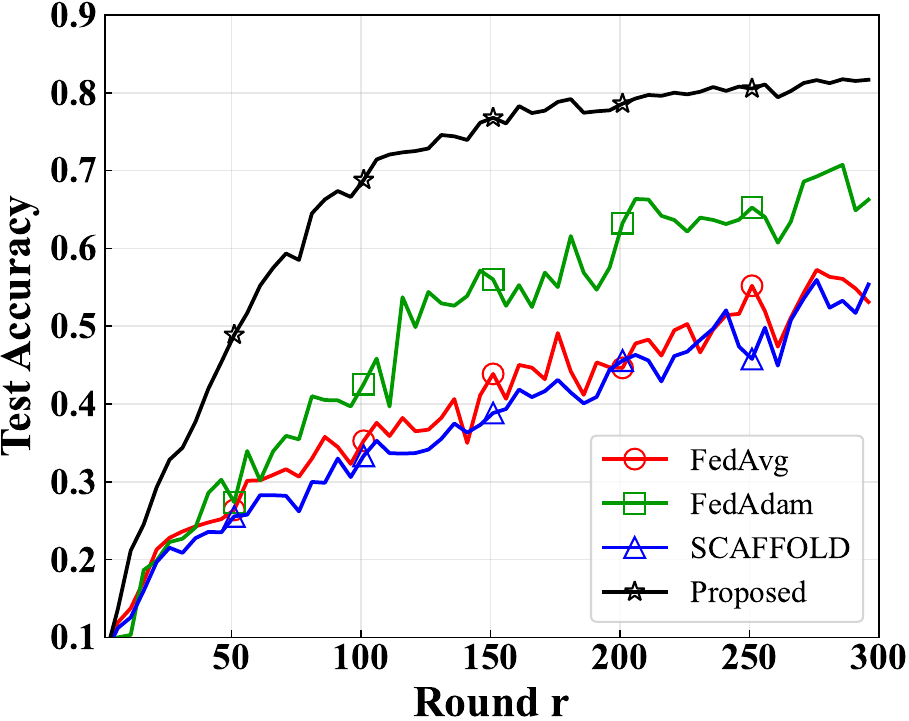}}
\subfigure[$dir=0.2$]{\includegraphics[width=0.44\textwidth]{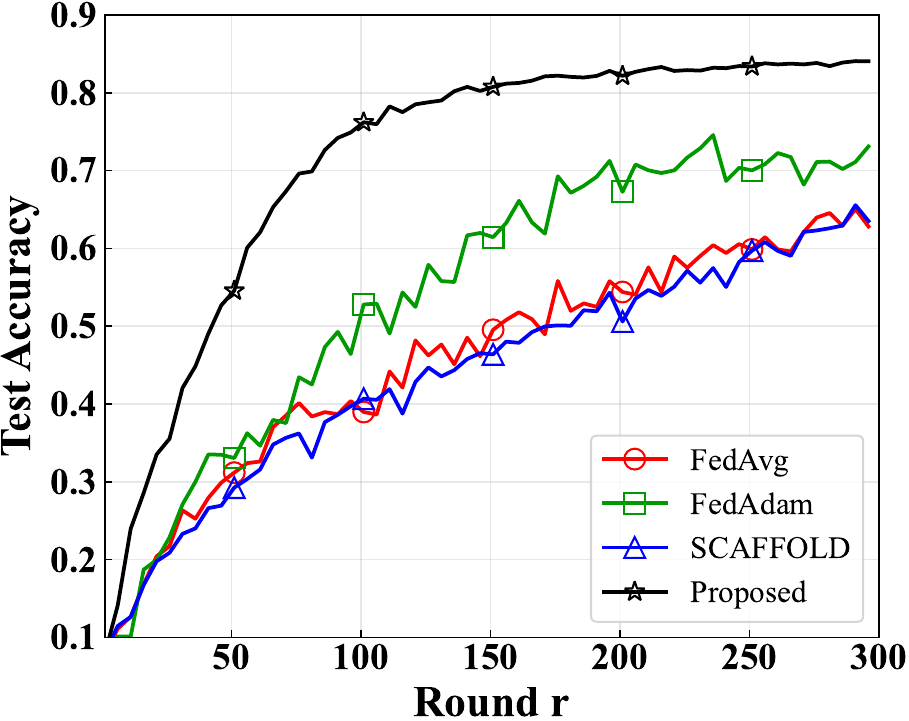}}\\
\vspace{-0.3cm}
\subfigure[$dir=0.3$]{\includegraphics[width=0.44\textwidth]{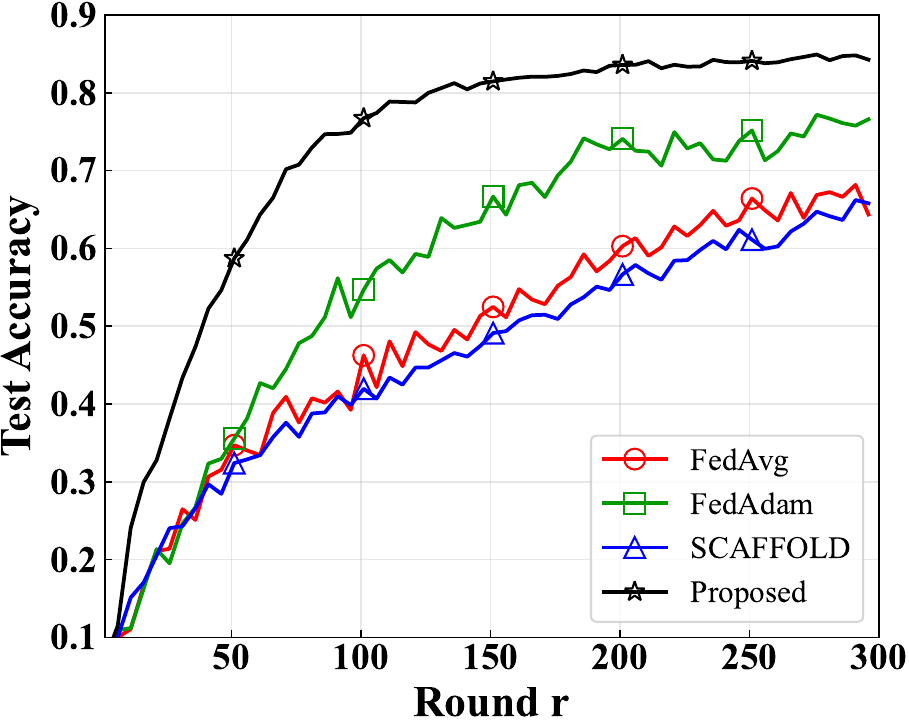}}
\subfigure[$dir=0.4$]{\includegraphics[width=0.44\textwidth]{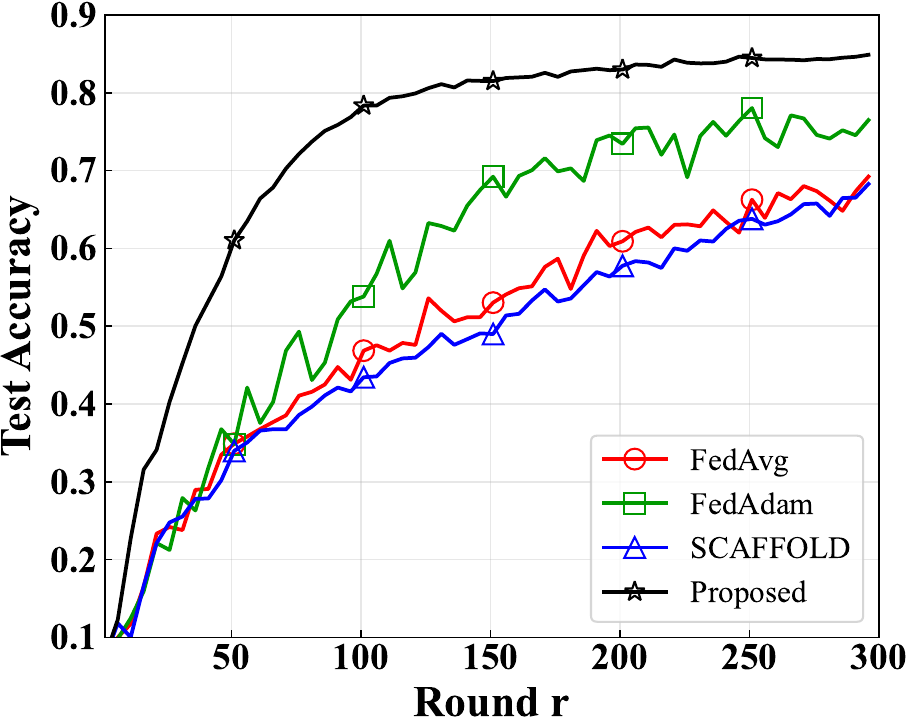}}
\vspace{-0.3cm}
\caption{Performance under different non-i.i.d. degrees.}
\label{fig:noniid_comparison}
\end{figure}

\vspace{-0.4cm}
\subsection{Impact of Dual-Quantization under \textbf{Fixed-QB}}

We first evaluate the proposed method against all baselines under \textbf{FIX-QB} with different choices of quantization levels. Fig. 1 presents their convergence performance when $B_i^r \in \{2,4\}$ and $B_D^r \in \{6,7,8\}$. As seen, under all tested quantization settings, the proposed method consistently achieves the fastest convergence and the highest accuracy. FedAvg and FedAdam suffer performance loss under low-bit transmission, while SCAFFOLD is less stable and also degrades in several low-bit cases. Increasing the uplink quantization bit generally improves all methods more noticeably than increasing the downlink bit, indicating the dominant role of uplink precision in aggregation quality. By contrast, the proposed method remains robust across different quantization settings and shows much smaller performance variation, especially in low quantization bit regimes.

\subsection{Impact of Non-i.i.d Data under \textbf{Fixed-QB}}

Fig.~\ref{fig:noniid_comparison} shows the convergence performance of all algorithms under test with the same quantization level ($B_i^r= 2$ and $B_D^r= 7$) but different choices of the Dirichlet parameter $dir$. Across all values of $dir \in \{0.1, 0.2, 0.3, 0.4\}$, the proposed method performs the best and significantly outperforms the others, demonstrating its strong robustness to data heterogeneity. Its advantage is particularly pronounced when $dir = 0.1$, where the non-i.i.d. effect is most severe. In contrast, FedAvg and FedAdam are more strongly affected by the heterogeneous data distribution, while SCAFFOLD performs better in some rounds but remains less stable. As $dir$ increases from $0.1$ to $0.4$, the performance of all methods improves, indicating that milder heterogeneity leads to easier FL. Nevertheless, the proposed method maintains a clear performance advantage throughout, confirming its effectiveness under fixed quantization and non-i.i.d. data.

\vspace{-0.1cm}
\subsection{Performance Evaluation under \textbf{SWAN-OPT}} 
\begin{figure}[!t]
\centering
\subfigure[FedAvg]{\includegraphics[width=0.23\textwidth]{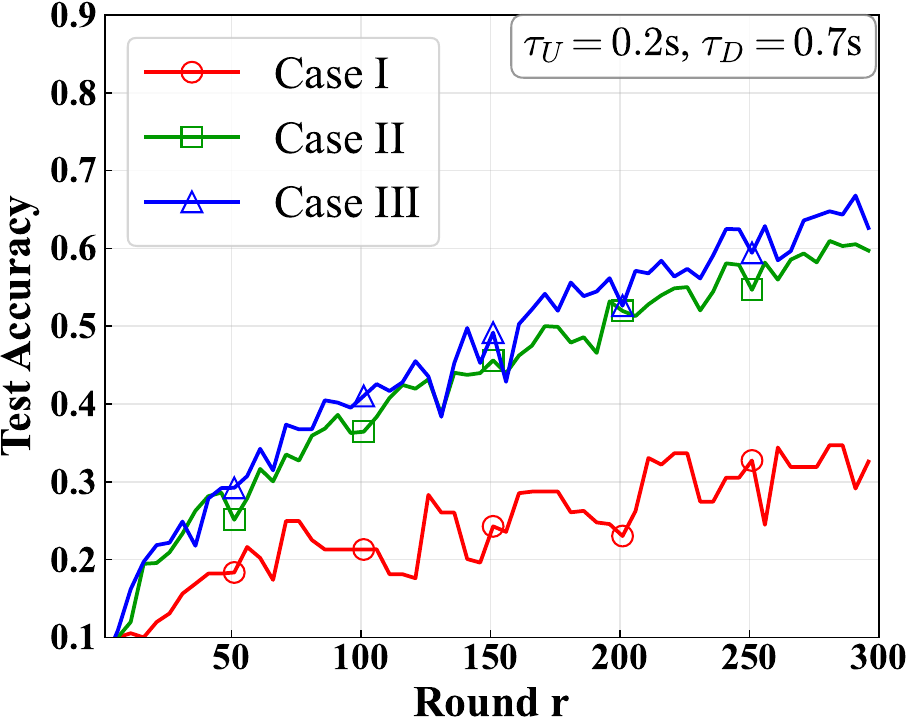}\label{fig:combined_swan_cases1_fedavg}}\hfill
\subfigure[SCAFFOLD]{\includegraphics[width=0.23\textwidth]{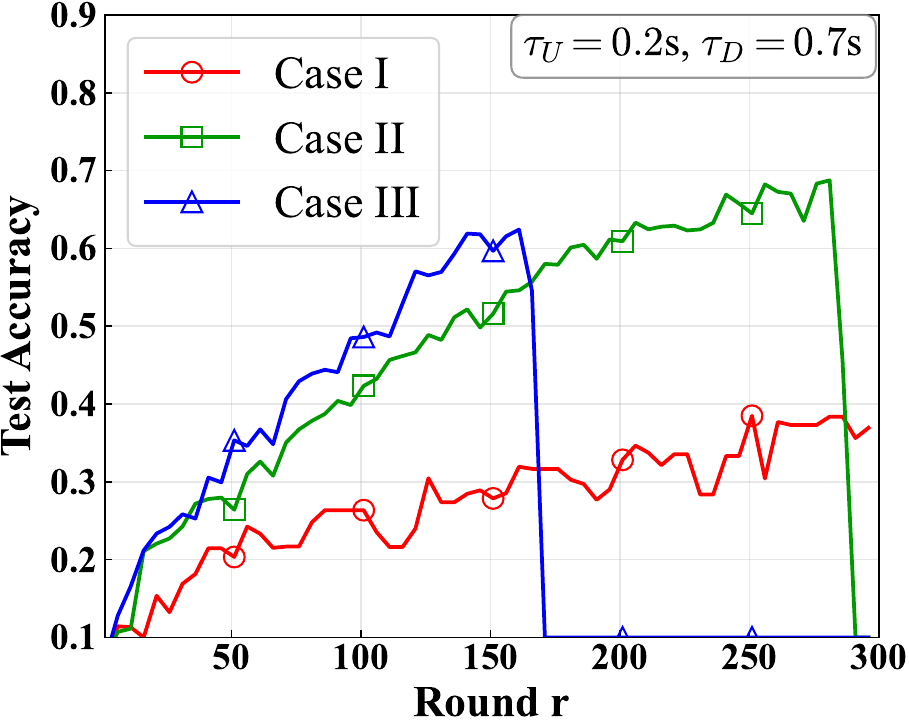}\label{fig:combined_swan_cases1_scaffold}}\hfill
\subfigure[FedAdam]{\includegraphics[width=0.23\textwidth]{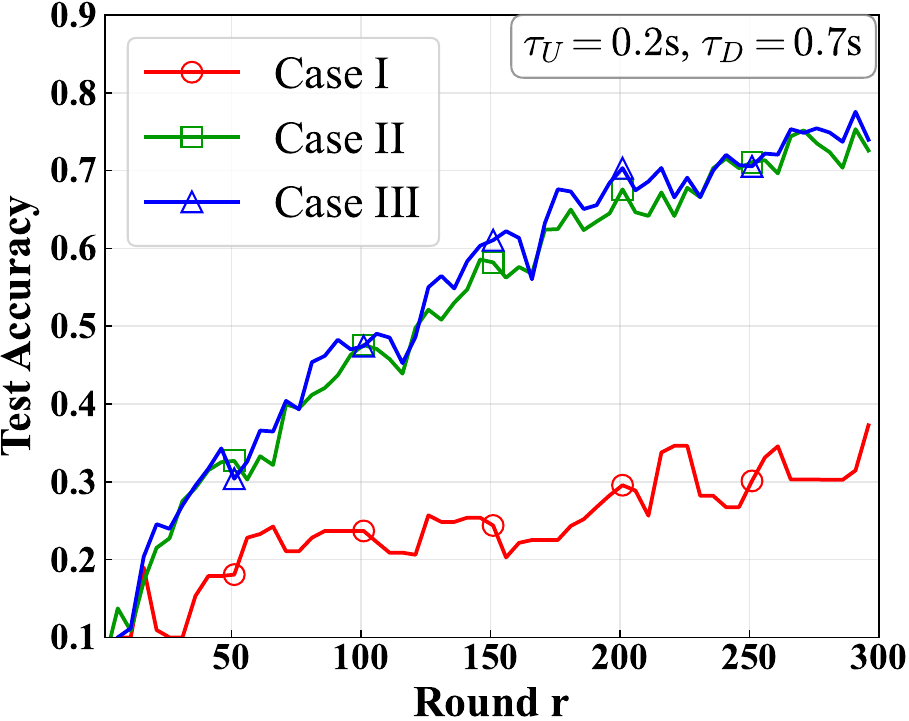}\label{fig:combined_swan_cases1_fedadam}}\hfill
\subfigure[Proposed]{\includegraphics[width=0.23\textwidth]{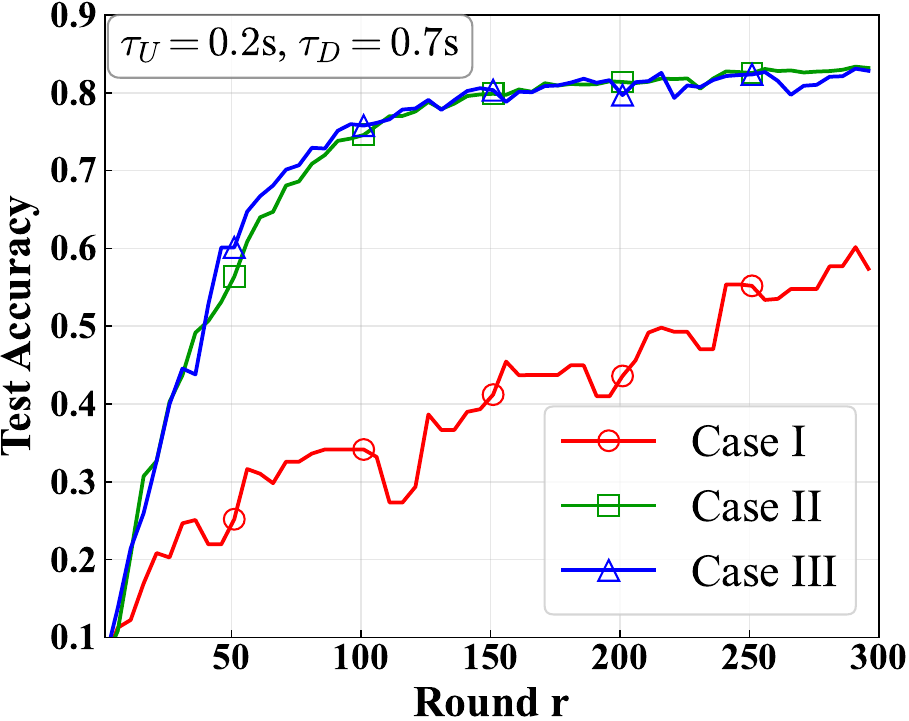}\label{fig:combined_swan_cases1_proposed}}\\
\vspace{-0.3cm}
\subfigure[FedAvg]{\includegraphics[width=0.23\textwidth]{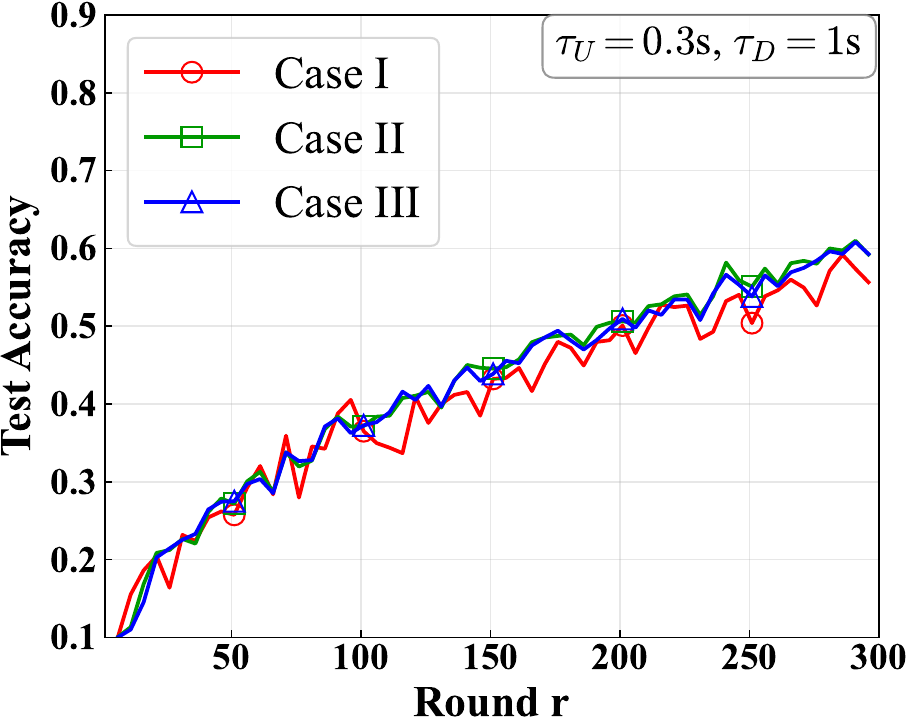}\label{fig:combined_swan_cases2_fedavg}}\hfill
\subfigure[SCAFFOLD]{\includegraphics[width=0.23\textwidth]{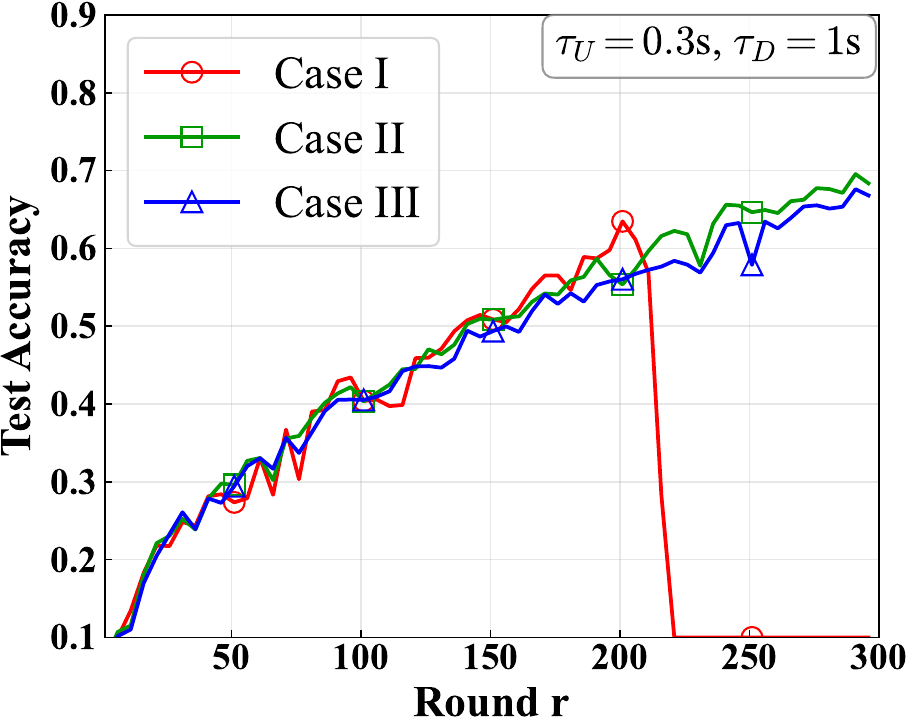}\label{fig:combined_swan_cases2_scaffold}}\hfill
\subfigure[FedAdam]{\includegraphics[width=0.23\textwidth]{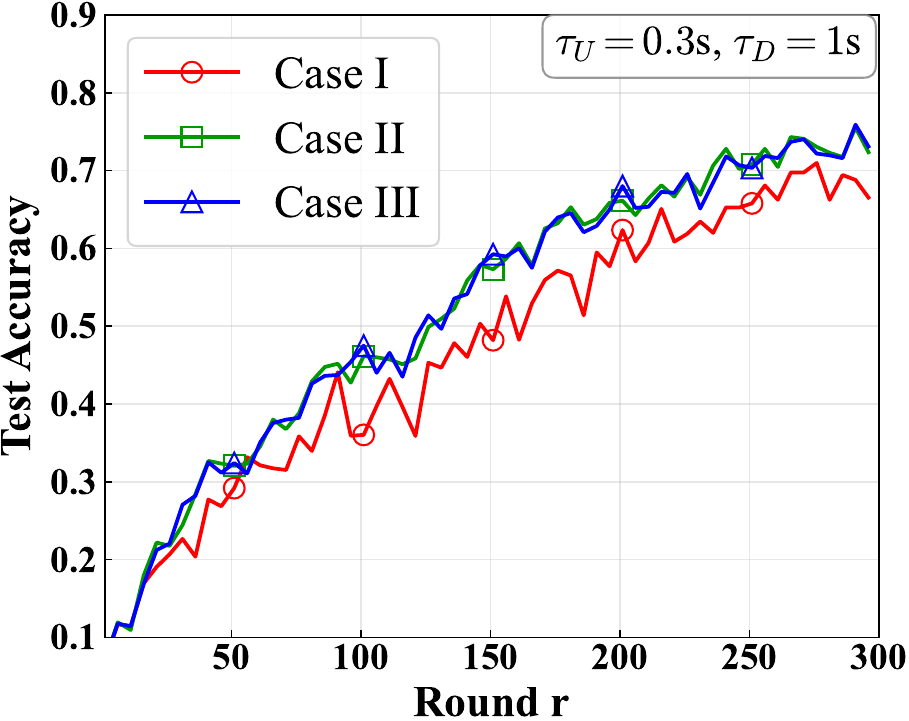}\label{fig:combined_swan_cases2_fedadam}}\hfill
\subfigure[Proposed]{\includegraphics[width=0.23\textwidth]
{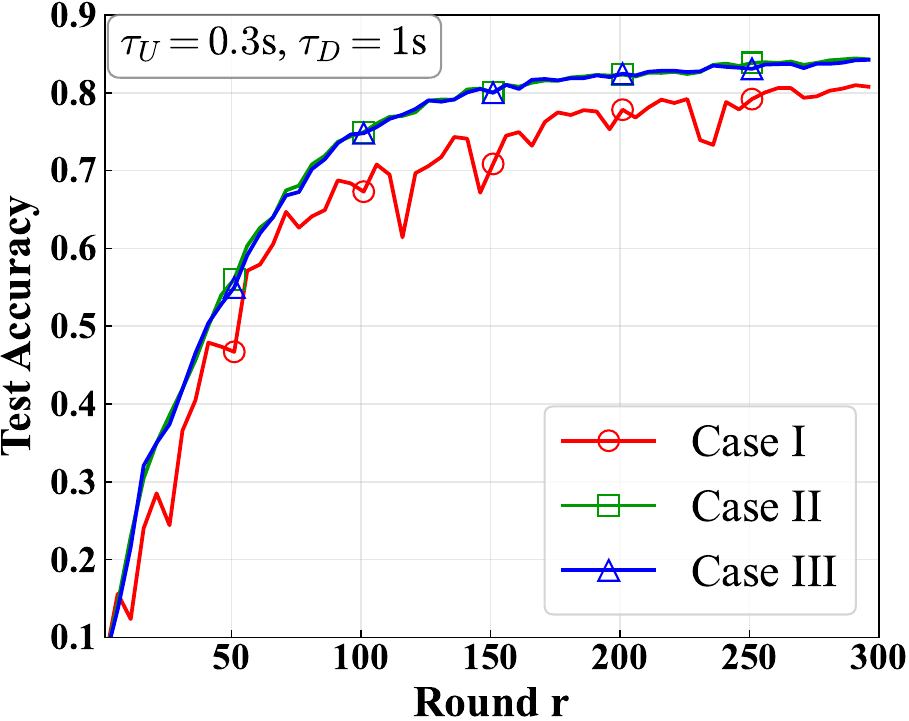}\label{fig:combined_swan_cases2_proposed}}\\
\vspace{-0.1cm}
\caption{Performance comparison under \textbf{SWAN-OPT} with different uplink designs and latency budgets.}
\label{fig:swan_cases_combined}
\end{figure}

We further evaluate the proposed method and all baselines under the \textbf{SWAN-OPT} configuration by considering the three uplink designs developed in Sec.~V. Specifically, Case I performs PA positioning under equal bandwidth allocation and fixed uplink quantization, Case II further enables client-adaptive uplink quantization under equal bandwidth allocation, and Case III jointly optimizes bandwidth allocation, PA positioning, and uplink quantization. The same SWAN-assisted downlink design is adopted in all three cases.

The results are shown in Fig. \ref{fig:swan_cases_combined}, where each row reports the performance of all compared methods under one specific pair of uplink and downlink latency budgets. In the first row, the latency budgets are stringent, with $\tau_U=0.2$s and $\tau_D=0.7$s. Under this setting, one can observe that each method with Case III generally outperforms its counterparts with Case II and Case I. Moreover, the performance gap between Case III and Case II is relatively small, whereas both of them exhibit a more noticeable advantage over Case I.  Besides, in the second row, where the latency budgets are relaxed to $\tau_U=0.3$s and $\tau_D=1$s, the performance gap among the three cases becomes much smaller. The curves of Case II and III are nearly overlapping, and both remain only slightly better than Case I.

Overall, Fig. \ref{fig:swan_cases_combined} shows that the benefit of uplink optimization is more evident under stringent latency constraints. In the stringent-latency regime, client-adaptive quantization in Case II already captures a large portion of the achievable gain, while the fully joint optimization in Case III provides further refinement through bandwidth allocation and PA positioning. When the latency constraint is less restrictive, the marginal gain brought by more sophisticated uplink optimization in Case II and III is reduced. We also remark that, across both latency settings, the proposed method consistently performs the best among all compared methods, confirming the effectiveness of integrating heterogeneity-robust FL
with SWAN-assisted communication optimization.
\vspace{-0.1cm}

\subsection{Comparison Between \textbf{SWAN-OPT} and \textbf{BS-OPT}}

\begin{figure}[!h]
\centering
\subfigure[Proposed vs SCAFFOLD]{\includegraphics[width=0.49\textwidth]{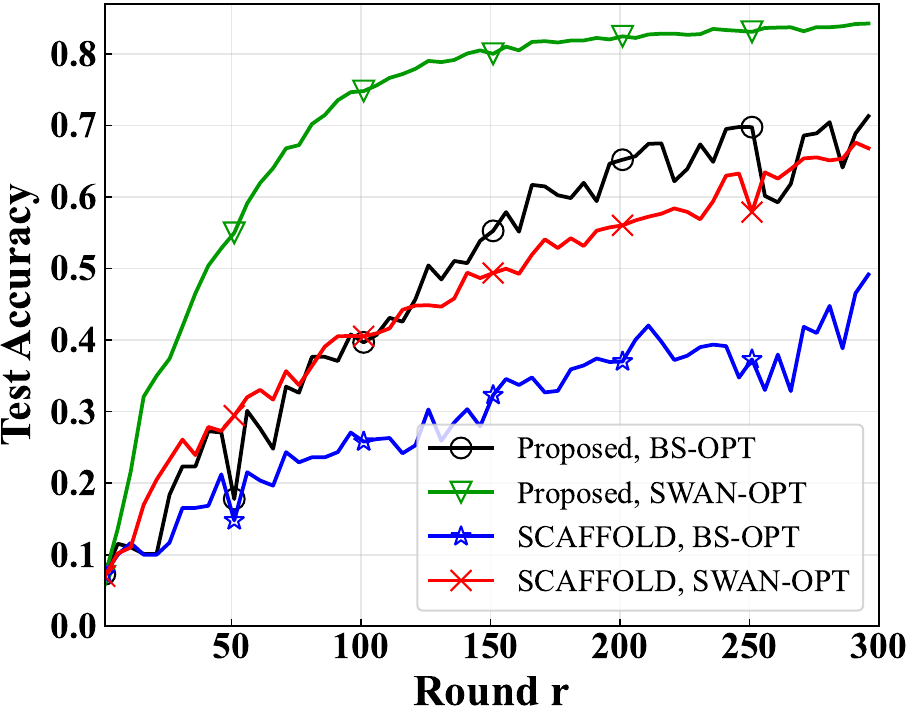}\label{fig:bs_swan_comparison_scaffold}}
\vspace{-0.1cm}
\subfigure[Proposed vs FedAdam]{\includegraphics[width=0.49\textwidth]{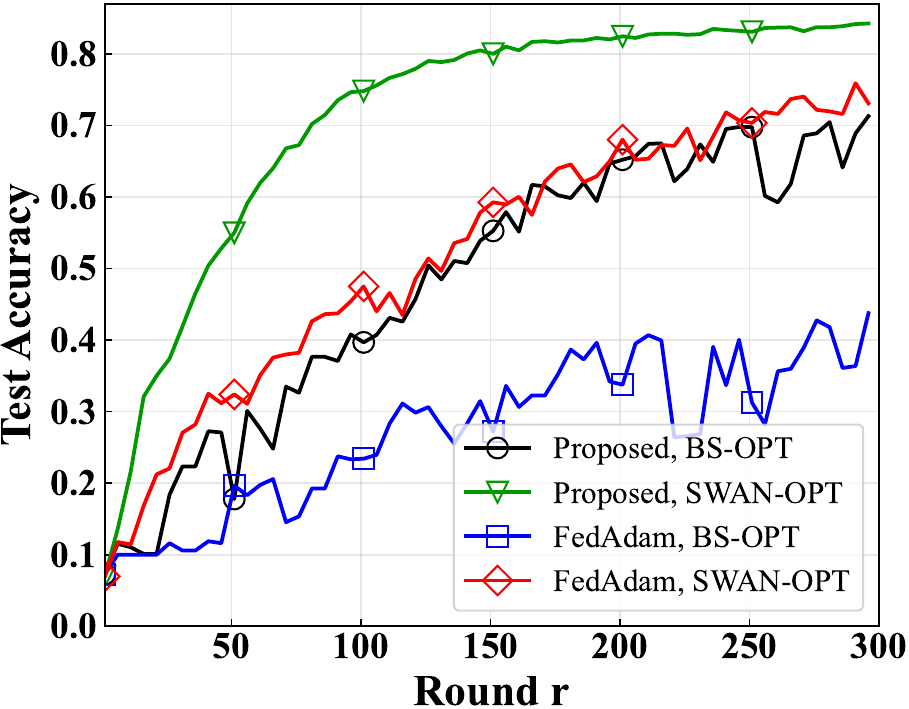}\label{fig:bs_swan_comparison_adam}}
\vspace{-0.1cm}
\caption{Comparison under \textbf{BS-OPT} and \textbf{SWAN-OPT}.}
\label{fig:bs_swan_comparison}
\end{figure}

Finally, we compare the proposed \textbf{SWAN-OPT} design with the conventional \textbf{BS-OPT} configuration under the setting of $\tau_U=0.3$s and $\tau_D=1$s. For \textbf{SWAN-OPT}, the uplink transmission in all methods adopts Case III. As shown in Fig. \ref{fig:bs_swan_comparison}, all compared methods achieve faster convergence and higher test accuracy under \textbf{SWAN-OPT} than under \textbf{BS-OPT}, demonstrating the advantage of SWAN in communication enhancement. More importantly, the proposed method remains clearly superior under both communication architectures, indicating that the proposed design can more effectively exploit the communication gains enabled by SWAN.

Table I further summarizes the detailed results under \textbf{FIX-QB}, \textbf{BS-OPT}, and \textbf{SWAN-OPT}. As seen, the proposed method achieves the best performance in all three scenarios. Under \textbf{SWAN-OPT}, it obtains the highest final accuracy of $84.18\%$, whereas the best baseline, FedAdam, achieves $76.26\%$. This corresponds to a relative accuracy improvement of about $10\%$. In addition, to reach $50\%, 60\%, 70\%$ accuracy, the proposed method requires only $40, 61, 81$ rounds, respectively, while FedAdam requires $114,146,225$ rounds. Hence, the proposed method is about $2.85\times$, $2.40\times$, and $2.77\times$ faster at these three target accuracy levels. Compared with its BS-based counterpart, the proposed SWAN-assisted design further improves the final accuracy by about $18.65\%$, while requiring only about half as many communication rounds to reach the same target accuracy. These results confirm that the performance gain comes from the joint effect of heterogeneity-robust learning and SWAN-enabled communication optimization, and further verify the effectiveness of PAs in boosting wireless FL.

\vspace{-0.4cm}

\section{Conclusion}\label{sec: conclusion}

\begin{table}[t!]
\centering
\setlength{\tabcolsep}{2.5pt}
\renewcommand{\arraystretch}{1.0}
\caption{}
\vspace{-0.3cm}
\label{tab:result_summary}
\parbox{0.98\columnwidth}{\small \#R-XX denotes the number of rounds for achieving XX\% accuracy. The symbol "-" means the algorithm crashes.}
\par\smallskip
\centering
\resizebox{1.0\columnwidth}{!}{\begin{tabular}{c| c| c| c| c| c| c}
\toprule
\midrule
\rowcolor{gray!20}
Scenario & Algorithm & ACC(\%) & \#R-50 ($\downarrow$) & \#R-60 ($\downarrow$) & \#R-70 ($\downarrow$) & \#R-80 ($\downarrow$)\\
\midrule
\multirow[c]{4}{*}{FIX-QB} & FedAvg & 65.51 & 99 & 154 & 246 & $>$300 \\
 & FedAdam & 75.32 & 79 & 98 & 141 & $>$300 \\
 & SCAFFOLD & 10.00 & 79 & 103 & - & - \\
 & \textbf{Proposed} & \textbf{82.98} & \textbf{28} & \textbf{38} & \textbf{57} & \textbf{131} \\
\midrule
\multirow{4}{*}{BS-OPT} & FedAvg & 45.15 & $>$300 & $>$300 & $>$300 & $>$300 \\
 & FedAdam & 44.28 & $>$300 & $>$300 & $>$300 & $>$300 \\
 & SCAFFOLD & 50.30 & 297 & $>$300 & $>$300 & $>$300 \\
 & \textbf{Proposed} & \textbf{70.95} & \textbf{126} & \textbf{166} & \textbf{241} & \textbf{$>$300} \\
\midrule
\multirow{4}{*}{SWAN-OPT} & FedAvg & 61.67 & 169 & 280 & $>$300 & $>$300 \\
 & FedAdam & 76.26 & 114 & 146 & 225 & $>$300 \\
 & SCAFFOLD & 68.33 & 133 & 194 & $>$300 & $>$300 \\
 & \textbf{Proposed} & \textbf{84.18} & \textbf{40} & \textbf{61} & \textbf{81} & \textbf{139} \\
\midrule
\end{tabular}
}
\vspace{-0.2cm}
\end{table}

This paper presents a PA-empowered FL framework. By integrating client-variance reduction and dual quantization with the SWAN architecture, the proposed method jointly optimizes beamforming, PA positioning, resource allocation, and quantization precision, thereby addressing both data heterogeneity and communication bottlenecks. In particular, a two-stage SWAN-assisted downlink protocol and a fairness-oriented uplink design are developed to support efficient model delivery and local-update transmission under practical wireless constraints. Experimental results demonstrate that the proposed framework consistently improves convergence speed and final model accuracy, highlighting the potential of PAs for robust and efficient wireless FL in heterogeneous networks.

\FloatBarrier
\appendices

\section{Proof of Proposition~\ref{prop:downlink_time_bit_recovery}}
\label{app:proof_prop_downlink_time_bit_recovery}

For a fixed admissible $B_D$ and positive rates, the two transmission constraints require
\begin{align}
\tau^{\rm I}&\geq \frac{d(B_D+1)+\mu}{R_{\rm mc}}, \notag\\
\tau_i^{\rm II}&\geq \frac{d(B_D+1)+\mu+b_{\rm ctl}}{R_{{\rm uc},i}},\quad i\in\Sc^r. \notag
\end{align}
The lower bounds are simultaneously attainable, giving the minimum required stage durations in \eqref{eq:downlink_tauI_star} and \eqref{eq:downlink_tauII_star}. Hence, $B_D$ is supportable if and only if
\begin{align}
[d(B_D+1)+\mu]
\left(\frac{1}{R_{\rm mc}}+\sum_{i\in\Sc^r}\frac{1}{R_{{\rm uc},i}}\right)
+b_{\rm ctl}\sum_{i\in\Sc^r}\frac{1}{R_{{\rm uc},i}}
\leq \tau_D. \label{eq:app_downlink_latency}
\end{align}
Solving this inequality for an integer $B_D$ yields the floor in \eqref{eq:downlink_bit_recovery}. If this value is below $B_D^{\min}$, no admissible precision is supportable for the given rates. Otherwise, it is the maximum admissible precision, and the minimum required durations provide one feasible time allocation; additional unused time can make other allocations feasible as well.

For any fixed admissible $B_D$, the left-hand side of \eqref{eq:app_downlink_latency} is nonincreasing in each rate. Because the multicast and unicast designs have independent feasible sets, their individual global maximizers can be used simultaneously. The resulting rates therefore support every $B_D$ achievable by any other design. If their recovered value satisfies $B_D^\star\geq B_D^{\min}$, it is globally optimal for \eqref{prob:two_phase_downlink}; otherwise, no design can meet the minimum precision and the problem is infeasible. \hfill$\blacksquare$

\section{Proof of Theorem~\ref{thm:2d_pa}}
\label{app:proof_thm1}

For $i\in\Uc_m^r$, let $\tilde g_i(x)=\rho_i(x)/\Gamma_i=((x-x_i)^2+c_i)/\Gamma_i$. Remove duplicate functions, which do not change their upper envelope $\tilde G_m(x)=\max_i\tilde g_i(x)$. The envelope is continuous and attains a minimum on the compact interval $\Lc_m$.

Let $x^\star$ be a minimizer. If it is a boundary point, it belongs to $\Cc_m^{\rm bdry}$. Otherwise, if two distinct functions are active at $x^\star$, they are equal there, so $x^\star\in\Cc_m^{\rm pair}$. This holds whether or not the envelope is differentiable at the intersection. If exactly one function $\tilde g_i$ is active, it remains uniquely active in a neighborhood of $x^\star$. First-order optimality then gives
\begin{align}
\tilde g_i'(x^\star)=2(x^\star-x_i)/\Gamma_i=0, \notag
\end{align}
and hence $x^\star=x_i\in\Cc_m^{\rm apex}$.

After duplicates are excluded, the difference of each pair of functions is a nonzero polynomial of degree at most two and has at most two real roots. Thus $\Cc_m$ is finite and contains every minimizer. Evaluating the envelope at its feasible candidates yields \eqref{eq:pa_opt}. Intuitively, an interior minimum occurs either at the bottom of one active quadratic or where active quadratics meet. \hfill$\blacksquare$

\section{Proof of Proposition~\ref{prop:case3_inverse_bandwidth}}
\label{app:proof_case3_inverse_bandwidth}

Fix an admissible $\bar B_U$ and PA positions. If $S(\bar B_U)<\gamma_{i,m}(x_m^{\rm PA})/\ln 2$, strict monotonicity of $C$ in bandwidth makes the rate requirement equivalent to
\begin{align}
W_i\geq C^{-1}\!\left(S(\bar B_U),\gamma_{i,m}(x_m^{\rm PA})\right). \notag
\end{align}
If the required rate is at least $\gamma_{i,m}(x_m^{\rm PA})/\ln 2$, no finite bandwidth can meet it, consistent with the value $+\infty$ assigned to the inverse.

Thus every feasible allocation has total bandwidth at least the sum of the inverse terms for its PA positions, which in turn is at least $G(\bar B_U)$. The budget therefore implies $G(\bar B_U)\leq W_U$.

Conversely, suppose $G(\bar B_U)\leq W_U$. Each segment objective in \eqref{eqn: min_bandwidth} is lower semicontinuous on the compact interval $\Lc_m$: its finite inverse terms are continuous and diverge as the required rate approaches the limiting rate. Hence each minimum is attained; it is finite under the stated inequality. Choose these segment minimizers and assign the corresponding inverse bandwidths. All rate constraints then hold with equality, the bandwidths are positive, and their sum is $G(\bar B_U)\leq W_U$. Empty segments have no bandwidth cost and their PA positions can be chosen arbitrarily. This proves the stated feasibility equivalence.

The equivalence gives \eqref{prob:inverse_prob} by projecting the feasible set onto $\bar B_U$. Because $S$ and the extended inverse are nondecreasing in the target, so is $G$. When the problem is feasible, choosing segment minimizers and inverse bandwidths at an optimal $\bar B_U^\star$ gives \eqref{eq:case3_opt_pa}--\eqref{eq:case3_opt_bw}, and the feasibility equivalence rules out any larger common floor. \hfill$\blacksquare$

\begingroup
\linespread{1.35}\selectfont
\smaller[1]
\bibliographystyle{IEEEtran}	
\bibliography{reference}

@article{li2025differentially,
  title = {Differentially Private Federated Stochastic Primal-Dual Learning for Internet of Vehicles},
  author = {Li, Yiwei and Wang, Shuai and Chang, Tsung-Hui},
  journal = {IEEE Internet Things J.},
  volume = {12},
  number = {11},
  pages = {17034-17050},
  year = {2025},
  doi = {10.1109/jiot.2025.3534982},
  month = jun
}

@inproceedings{FedVRA_Shuai_2023,
  title = {Beyond {ADMM}: A Unified Client-Variance-Reduced Adaptive Federated Learning Framework},
  booktitle = {Proc. AAAI},
  year = {2023},
  pages = {10175-10183},
  doi = {10.1609/aaai.v37i8.26212},
  author = {Shuai Wang and Yanqing Xu and Zhiguo Wang and Tsung-Hui Chang and Tony Q. S. Quek and Defeng Sun},
  address = {Washington, DC, USA},
  month = {Feb. 7--14}
}

@inproceedings{AdaptiveFL_2021,
  title = {Adaptive Federated Optimization},
  booktitle = {Proc. ICLR},
  address = {Virtual},
  year = {2021},
  author = {Sashank Reddi and others},
  month = {May 3--7}
}

@ARTICLE{IterativeMMSE,
  author = {Shi, Qingjiang and Razaviyayn, Meisam and Luo, Zhi-Quan and He, Chen},
  journal = {IEEE Trans. Signal Process.},
  year = {2011},
  volume = {59},
  number = {9},
  pages = {4331-4340},
  doi = {10.1109/tsp.2011.2147784},
  month = sep,
  title = {An Iteratively Weighted {MMSE} Approach to Distributed Sum-Utility Maximization for a {MIMO} Interfering Broadcast Channel}
}

@ARTICLE{SDRQOP_2010,
  author = {Luo, Zhi-Quan and Ma, Wing-Kin and So, Anthony Man-Cho and Ye, Yinyu and Zhang, Shuzhong},
  journal = {IEEE Signal Process. Mag.},
  title = {Semidefinite Relaxation of Quadratic Optimization Problems},
  year = {2010},
  volume = {27},
  number = {3},
  pages = {20-34},
  doi = {10.1109/msp.2010.936019},
  month = may
}

@ARTICLE{MBO_pinching,
  author = {Wang, Zhaolin and Ouyang, Chongjun and Mu, Xidong and Liu, Yuanwei and Ding, Zhiguo},
  journal = {IEEE Trans. Commun.},
  title = {Modeling and Beamforming Optimization for Pinching-Antenna Systems},
  year = {2025},
  volume = {73},
  number = {12},
  pages = {13904-13919},
  doi = {10.1109/tcomm.2025.3621049},
  month = dec
}

@inproceedings{FedAvg_noniid_2019,
  author = {X. Li and K. Huang and W. Yang and S. Wang and Z. Zhang},
  booktitle = {Proc. ICLR},
  address = {Virtual},
  year = {2020},
  month = {Apr. 26--May 1},
  title = {On the Convergence of {FedAvg} on {Non-IID} Data}
}

@ARTICLE{FedQVR_Shuai_2025,
  author = {Shuai Wang and Yanqing Xu and Chaoqun You and Mingjie Shao and Tony Q. S. Quek},
  journal = {IEEE Trans. Mobile Comput.},
  title = {Communication-Efficient Federated Learning by Quantized Variance Reduction for Heterogeneous Wireless Edge Networks},
  year = {2025},
  volume = {24},
  number = {8},
  pages = {7632-7647},
  doi = {10.1109/tmc.2025.3551759},
  month = aug
}

@ARTICLE{SWAN-2026,
  author = {Ouyang, Chongjun and Jiang, Hao and Wang, Zhaolin and Liu, Yuanwei and Ding, Zhiguo},
  journal = {IEEE Trans. Commun.},
  title = {Uplink and Downlink Communications in Segmented Waveguide-Enabled Pinching-Antenna Systems {(SWANs)}},
  year = {2026},
  volume = {74},
  number = {},
  pages = {3688-3703},
  doi = {10.1109/tcomm.2026.3653885},
}

@Book{BK:BoydV04,
  author = {S. Boyd and L. Vandenberghe},
  title = {Convex {O}ptimization},
  publisher = {Cambridge University Press},
  year = {2004},
  doi = {10.1017/cbo9780511804441},
  address = {Cambridge, U.K.}
}

@inproceedings{karimireddy2020scaffold,
  author = {Karimireddy, Sai Praneeth and Kale, Satyen and Mohri, Mehryar and Reddi, Sashank and Stich, Sebastian and Suresh, Ananda Theertha},
  booktitle = {Proc. ICML},
  address = {Virtual},
  year = {2020},
  pages = {5132-5143},
  title = {{SCAFFOLD}: Stochastic Controlled Averaging for Federated Learning},
  month = {Jul. 13--18},
  volume = {119},
  publisher = {PMLR}
}

@article{wang2025robust,
  author = {Wang, Yanmeng and Ji, Wenkai and Zhou, Jian and Xiao, Fu and Chang, Tsung-Hui},
  journal = {IEEE Trans. Mobile Comput.},
  title = {Robust Federated Learning in Unreliable Wireless Networks: A Client Selection Approach},
  year = {2026},
  volume = {25},
  number = {4},
  pages = {5181-5198},
  doi = {10.1109/tmc.2025.3628154},
  month = apr
}

@article{wang2022quantized,
  title = {Quantized federated learning under transmission delay and outage constraints},
  author = {Wang, Yanmeng and Xu, Yanqing and Shi, Qingjiang and Chang, Tsung-Hui},
  journal = {IEEE J. Sel. Areas Commun.},
  volume = {40},
  number = {1},
  pages = {323--341},
  year = {2022},
  doi = {10.1109/jsac.2021.3126081},
  month = jan
}

@ARTICLE{Straggler-FL-pinn2026,
  author = {Wu, Bibo and Fang, Fang and Zeng, Ming and Wang, Xianbin},
  journal = {IEEE Trans. Veh. Technol.},
  title = {Straggler-Resilient Federated Learning over A Hybrid Conventional and Pinching Antenna Network},
  year = {2026},
  doi = {10.1109/tvt.2026.3656768},
  month = jul,
  volume = {75},
  number = {7},
  pages = {15404--15409}
}

@ARTICLE{xu2025rate,
  author = {Xu, Yanqing and Ding, Zhiguo and Karagiannidis, George K.},
  journal = {IEEE Wireless Commun. Lett.},
  title = {Rate Maximization for Downlink Pinching-Antenna Systems},
  year = {2025},
  volume = {14},
  number = {5},
  pages = {1431-1435},
  doi = {10.1109/lwc.2025.3543889},
  month = may
}

@ARTICLE{xu2026attenuation,
  author = {Xu, Yanqing and Ding, Zhiguo and Schober, Robert and Chang, Tsung-Hui},
  journal = {IEEE Trans. Wireless Commun.},
  title = {Pinching-Antenna Systems With In-Waveguide Attenuation: Performance Analysis and Algorithm Design},
  year = {2026},
  volume = {25},
  number = {},
  pages = {14564-14580},
  doi = {10.1109/twc.2026.3676036},
}

@ARTICLE{FL_MoblieEdge_2019,
  title = {Federated Learning in Mobile Edge Networks: A Comprehensive Survey},
  volume = {22},
  number = {3},
  pages = {2031--2063},
  year = {2020},
  doi = {10.1109/comst.2020.2986024},
  author = {Wei Yang Bryan Lim and others},
  journal = {IEEE Commun. Surveys Tuts.},
  month = {Third Quarter}
}

@ARTICLE{1634819,
  journal = {IEEE Trans. Signal Process.},
  title = {Transmit beamforming for physical-layer multicasting},
  year = {2006},
  volume = {54},
  number = {6},
  pages = {2239-2251},
  doi = {10.1109/tsp.2006.872578},
  month = jun,
  author = {Sidiropoulos, N. D. and Davidson, T. N. and Luo, Zhi-Quan}
}

@ARTICLE{5594708,
  author = {Gesbert, David and Hanly, Stephen and Huang, Howard and Shamai Shitz, Shlomo and Simeone, Osvaldo and Yu, Wei},
  journal = {IEEE J. Sel. Areas Commun.},
  title = {Multi-Cell {MIMO} Cooperative Networks: A New Look at Interference},
  year = {2010},
  volume = {28},
  number = {9},
  pages = {1380-1408},
  doi = {10.1109/jsac.2010.101202},
  month = dec
}

@ARTICLE{samy2025pass_ris_mmwave,
  author = {Samy, Mostafa and Al-Hraishawi, Hayder and Alsenwi, Madyan and Adam, Abuzar B. M. and Chatzinotas, Symeon and Ottersten, Björn},
  journal = {IEEE Wireless Commun. Lett.},
  title = {Pinching Antenna Systems Versus Reconfigurable Intelligent Surfaces in {mmWave}},
  year = {2025},
  volume = {14},
  number = {12},
  pages = {4022-4026},
  doi = {10.1109/lwc.2025.3611699},
  month = dec
}

@article{xu2025random_los_nlos_pa,
  author = {Yanqing Xu and Yang Lu and Zhiguo Ding and Tsung-Hui Chang},
  title = {Pinching-Antenna System Design under Random {LoS} and {NLoS} Channels},
  doi = {10.1109/twc.2026.3698089},
  journal = {IEEE Trans. Wireless Commun.},
  volume = {25},
  pages = {18440--18455},
  year = {2026},
}

@Article{e28030341,
  AUTHOR = {Lin, Yushen and Ding, Zhiguo},
  TITLE = {Tail-Latency-Aware Federated Learning with Pinching Antenna: Latency, Participation, and Placement},
  JOURNAL = {Entropy},
  VOLUME = {28},
  YEAR = {2026},
  NUMBER = {3},
  PubMedID = {41899993},
  ISSN = {1099-4300},
  doi = {10.3390/e28030341},
  month = mar,
  note = {Art. no. 341}
}
\endgroup

\end{document}